\documentclass[fleqn,usenatbib]{mnras}

\usepackage{newtxtext,newtxmath}

\usepackage[T1]{fontenc}

\DeclareRobustCommand{\VAN}[3]{#2}
\let\VANthebibliography\thebibliography
\def\thebibliography{\DeclareRobustCommand{\VAN}[3]{##3}\VANthebibliography}

\usepackage{graphicx}	
\usepackage{amsmath}	
\usepackage{float}
\usepackage{geometry}
\usepackage{booktabs}
\usepackage{lscape}
\usepackage{threeparttable}
\usepackage{xcolor}
\usepackage{ulem}
\usepackage{cancel}

\newcommand{\nickel}{\ensuremath{^{56}\mathrm{Ni}}}
\newcommand{\cobalt}{\ensuremath{^{56}\mathrm{Co}}}
\newcommand{\mch}{\ensuremath{M_\mathrm{Ch}}}
\newcommand{\artis}{\textsc{artis}}
\newcommand{\tilda}{$\sim $ }
\newcommand{\msun}{\ensuremath{\mathrm{M}_\odot}}

\title[Luminous Remnants for SNe~Iax]{
Including a Luminous Central Remnant in Radiative Transfer Simulations for Type Iax Supernovae}

\author[F. P. Callan et al.]{F. P. Callan,$^{1}$\thanks{E-mail: f.callan@qub.ac.uk}
S. A. Sim,$^{1}$ C. E. Collins,$^2$ L. J. Shingles,$^2$ F. Lach,$^{3}$ F. K. R\"opke,$^{3,4}$ R. Pakmor,$^5$ \newauthor M. Kromer$^{3}$ and S. Srivastav$^1$
\\
$^{1}$School of Maths and Physics, Queen's University Belfast, University Road, Belfast BT7 1NN, UK \\
$^{2}$GSI Helmholtzzentrum f\"{u}r Schwerionenforschung, Planckstraße 1, 64291 Darmstadt, Germany\\
$^3$Heidelberger Institut f\"ur Theoretische Studien, Schloss-Wolfsbrunnenweg 35, D-69118, Heidelberg, Germany\\
$^4$Zentrum f{\"u}r Astronomie der Universit{\"a}t Heidelberg, Institut f{\"u}r Theoretische Astrophysik, Philosophenweg 12, D-69120 Heidelberg, Germany\\
$^5$Max-Planck-Institut f\"{u}r Astrophysik, Karl-Schwarzschild-Str. 1, D-85748, Garching, Germany
}

\date{Accepted XXX. Received YYY; in original form ZZZ}

\pubyear{2023}

\begin{document}
\label{firstpage}
\pagerange{\pageref{firstpage}--\pageref{lastpage}}
\maketitle

\begin{abstract}
Type Iax supernovae (SNe~Iax) are proposed to arise from deflagrations of Chandrasekhar mass white dwarfs (WDs). Previous deflagration simulations have achieved good agreement with the light curves and spectra of intermediate-luminosity and bright SNe~Iax. However, the model light curves decline too quickly after peak, particularly in red optical and near-infrared (NIR) bands. Deflagration models with a variety of ignition configurations do not fully unbind the WD, leaving a remnant polluted with \nickel. Emission from such a remnant may contribute to the luminosity of SNe~Iax. Here we investigate the impact of adding a central energy source, assuming instantaneous powering by \nickel~decay in the remnant, in radiative transfer calculations of deflagration models. Including the remnant contribution improves agreement with the light curves of SNe~Iax, particularly due to the slower post-maximum decline of the models. Spectroscopic agreement is also improved, with intermediate-luminosity and faint models showing greatest improvement. We adopt the full remnant \nickel~mass predicted for bright models, but good agreement with intermediate-luminosity and faint SNe~Iax is only possible for remnant \nickel~masses significantly lower than those predicted. This may indicate that some of the \nickel~decay energy in the remnant does not contribute to the radiative luminosity but instead drives mass ejection, or that escape of energy from the remnant is significantly delayed. Future work should investigate the structure of remnants predicted by deflagration models and the potential roles of winds and delayed energy escape, as well as extend radiative transfer simulations to late times. 
      
\end{abstract}




\section{Introduction}
\label{sec:intro}
Through modern transient surveys it has become clear that Type Ia supernovae (SNe~Ia) are a diverse class of objects: in addition to the ‘‘normal'' SNe~Ia, which exhibit an empirical relation between their peak brightness and light curve evolution \citep{phillips1993a}, there are multiple distinct sub-classes (see e.g. \citealt{taubenberger2017a} for a review).

Type Iax supernovae (SNe~Iax, \citealt{li2003a, foley2013a}) are the most numerous peculiar  sub-class of SNe Ia, recently estimated by \cite{srivastav2022a} to make up $15^{+15}_{-9}\%$ of the total SNe Ia rate. Compared to normal SNe Ia, SNe~Iax show low absolute brightness at peak relative to their light curve widths (e.g. \citealt{mcclelland2010a, stritzinger2014a}) and thus fall outside the width-luminosity relation. They display a wide diversity in their band light curve properties showing variations in peak brightness of up to six magnitudes in certain bands and diversity in their light curve rise times and decline rates (\citealt{magee2016a, srivastav2022a}).  

SNe~Iax are spectroscopically distinct from normal SNe Ia: the blue continua and absorption from higher ionisation species observed in their early-time spectra mean they show more similarities to the sub-classes of 1991T-like and 1999a-like SNe Ia, which exhibit hot photospheres (\citealt{foley2013a, mccully2014b}). Additionally, SNe~Iax spectra show significantly lower expansion velocities at peak compared to normal SNe Ia (\citealt{foley2013a, stritzinger2015a}). The late time spectra of SNe~Iax show a large number of permitted lines from iron group elements (IGEs) along with some intermediate mass elements (IMEs) and potential signatures of oxygen, while forbidden lines of IGEs and IMEs are also present \citep{jha2006b, foley2016a}. As such, SNe~Iax never enter a fully nebular phase (unlike normal SNe Ia \citealt{graham2017a, taubenberger2015a}). The SNe~Iax sub-class also shows a large spectroscopic diversity in line velocities, strengths and widths particularly at later times (\citealt{mccully2014b, stritzinger2015a, yamanaka2015a, foley2016a}). 

Multiple explosion scenarios have been suggested to explain SNe~Iax, such as the core collapse of massive stars (\citealt{foley2009a, valenti2009a, moriya2010a}), pulsationally delayed detonations (\citealt{stritzinger2015a}), the failed detonation of an oxygen-neon (ONe) WD merging with a carbon-oxygen (CO) secondary \citep{kashyap2018a}, a WD-neutron star merger \citep{fernandez2013a, bobrick2022a} and a detonation in a helium shell surrounding a CO WD \citep{foley2009a}. 

However, amongst the most widely discussed scenarios is that in which SNe~Iax are pure deflagrations of Chandrasekhar mass (\ensuremath{M_\mathrm{Ch}}) CO WDs (\citealt{branch2004a, phillips2007a}), as the properties of such explosions naturally reproduce many of the observed characteristics of SNe~Iax. In particular, simulations of pure deflagrations in \mch~CO WDs \citep{jordan2012b, kromer2013a, kromer2015a, fink2014a, long2014a, leung2020a, lach2022a} reproduce many of the observed characteristics of SNe~Iax, such as lower \nickel~masses and hence sub-luminous light curves as well as lower expansion velocities. Furthermore, the pure deflagration model sequences presented by \cite{kromer2013a, fink2014a} and \cite{lach2022a} (hereafter K13, F14 and L22 respectively) are together able to reproduce almost the full diversity in peak brightness of observed SNe~Iax by varying the initial geometry of the explosion simulations, failing to match only the faintest observed SNe~Iax. These models also produce light curves and spectra in reasonably good agreement with bright and intermediate-luminosity SNe~Iax, particularly at earlier epochs. Additionally, \cite{camacho-neves2023a} have demonstrated good spectroscopic agreement between a deflagration model and the SN Iax, SN~2014dt, over \tilda 500 days of its evolution.

The near-infrared light curves of SNe~Iax further support the pure deflagration scenario: the lack of a prominent secondary maximum that is observed for normal SNe Ia (\citealt{li2003a}) points to a well-mixed ejecta structure for SNe~Iax (\citealt{jha2006b, phillips2007a, sahu2008a}), consistent with turbulent deflagration models (although see \citealt{stritzinger2015a} and \citealt{magee2022a} for further discussion of stratification in SNe~Iax). 

Despite the generally good agreement between \mch~CO WD pure deflagration models and observed SNe~Iax, some systematic differences remain. Specifically, the K13, F14 and L22 model light curves decline too quickly post peak, particularly in the red and NIR bands, meaning the bolometric light curves also decline too rapidly. This systematic difference worsens when comparing to fainter SNe~Iax where the decline post peak of the models is too fast compared to observed SNe~Iax in all bands. The faster decline for red, optical, and NIR wavelengths after peak relative to other bands also makes the model spectra too blue at later times relative to observed SNe~Iax. 

The outcome of pure deflagration simulations is sensitive to the initial choice of ignition conditions \citep{jordan2012b, kromer2013a, fink2014a, long2014a, kromer2015a, leung2020a, lach2022a}. For a wide variety of ignition configurations, the explosion is not energetic enough to fully unbind the WD \citep{jordan2012b, kromer2013a, fink2014a} potentially leaving behind a stellar remnant polluted with burning products from the deflagration, including \nickel. The remnant is therefore expected to be luminous. Such a luminous remnant may have been detected in late time observations (\tilda 4 years after explosion) of the faint SN~Iax, SN~2008ha \citep{foley2014a} and bright SN~Iax, SN 2012Z \citep{mccully2022a}. Further evidence of a remnant may be present in the slowly declining late time light curves of SN~2014dt \citep{kawabata2018a} and SN~2019muj \citep{kawabata2021a} and the peculiar late time spectra of SNe~Iax that never become truly nebular (\citealt{kromer2013a,foley2016a, kawabata2018a, kawabata2021a,maeda2022a, camacho-neves2023a}).

Accounting for the remnant may be key to fully understanding the observed properties of SNe~Iax. An exploratory calculation by K13 found adding the energy deposited by radioactive \nickel~in the remnant to the light curve predicted for their N5def deflagration model significantly improved the agreement with the bolometric light curve of SN~2005hk, a bright SNe~Iax, particularly at later times. However, their calculation simply adds this energy to the lightcurve computed from the ejecta model and does not take into account the interaction of this radiation with the ejecta. Therefore, with this approach only a crude bolometric light curve can be calculated and the early-time bolometric light curves will not be accurately represented. 

\cite{shen2017a} also found their models of post SNe~Iax remnant winds to be consistent with late time photometric observations of SNe~Iax. Additionally, from analysis of the late-time spectra of the intermediate-luminosity SNe~Iax, SN~2019muj, \cite{maeda2022a} find that a denser inner ejecta component is present. They attribute this to a secondary mass ejection from a remnant. The faint hybrid carbon-oxygen-neon (CONe) WD pure deflagration model which \cite{kromer2015a} produced as an attempt to match SN~2008ha also predicts a luminous remnant which they suggest may have a significant impact on the optical display of the model.

The aim of this paper is to incorporate a central source representing the remnant material into our 3D, time-dependent radiative transfer simulations in order to predict how a luminous remnant impacts the light curves and spectra predicted for \mch~CO WD pure deflagration models covering a range of peak luminosities. This improves on the exploratory calculation of K13 since we can simulate the early time diffusion of this radiation through the ejecta and investigate how it affects light curves, colours and spectra. We take the pure deflagration models presented by L22 as a starting point. Informed by the remnant compositions predicted by the  hydrodynamic explosion simulations of L22 we add a central \nickel~source representing a luminous remnant to a sub-set of the standard L22 deflagration models. We discuss the treatment we adopt for the remnant and summarise our radiative transfer simulation set up in Section \ref{sec:numerical_methods}. In Section \ref{sec:results} we make detailed comparisons between the light curves and spectra predicted by our radiative transfer simulations and those of observed SNe~Iax, before discussing the implications of our results along with our conclusions in Section \ref{sec:Discussion_Conclusions}.

\section{Numerical Methods}  
\label{sec:numerical_methods}
\subsection{Explosion models}
\label{subsec:explosion_simulations}
The sequence of \mch~CO WD pure deflagration models presented by L22 are the basis for this work. These models all adopt a single-spark ignition, which has been argued to be a more realistic ignition configuration than a multi-spark ignition as adopted by F14 (\citealt{kuhlen2006a, zingale2009a, zingale2011a, nonaka2012a}). The L22 sequence provides a good starting point to explore the impact of including a luminous remnant in our simulations: all models in this sequence produce explosions that are not energetic enough to fully unbind the WD, leaving behind remnants with masses greater than a solar mass that are polluted with burning products from the deflagration including \nickel. Additionally, the L22 model sequence covers almost the full peak brightness range of the SNe~Iax class, failing to reproduce only the very brightest and faintest observed SNe~Iax. We can therefore explore how a luminous remnant impacts our model comparisons with a wide variety of observed SNe~Iax from bright to faint events. 

The \textsc{leafs} (\citealt{reinecke1999b}) hydrodynamic explosion simulations used to calculate the L22 models adopted an expanding grid to track the ejecta \citep{roepke2005a}. This means the remnant is not well resolved at the end of the hydrodynamic explosion simulations (at $t \, {=}\, 100 \, \mathrm{s}$). Therefore, only basic properties of the remnant can be extracted: the overall mass and composition at the end of the explosion simulations can be determined from the tracer particles which remain bound (see L22 for details). We note however, that while we can say the remnant is bound at $t \,{=}\, 100\,\mathrm{s}$, this does not necessarily mean that the material in the remnant will all remain bound at later times. Indeed, the energy predicted to be deposited in the remnants from the \nickel~decay chain indicates they are super-Eddington by a factor of ${\sim}\,1000$ to $10000$ depending on the model and time in the simulation. This means it is highly likely that some or all of the remnant material becomes unbound, perhaps in the form of a radioactively driven post supernova wind as proposed by \citet{shen2017a} or as a secondary mass ejection as described by \cite{maeda2022a}. Therefore, when we discuss the remnant in this work we are referring to any material which remains bound up to the end of the hydrodynamic explosion simulation at 100\,s and so would not be included as part of the ejecta in the radiative transfer simulations for the standard L22 model sequence. 

The remnants' structure at the end of the L22 explosion simulations is that of a puffed-up envelope of burning products from the deflagration, mixed with CO material, which is settling onto a relatively dense CO core (see also K13 and \citealt{jordan2012b}). The remnant contains a significant fraction of \nickel~(between 2 and 7\% depending on the model) with this \nickel~predicted to be primarily found in the envelope (again see L22 for details). For the models in this work, we calculated an estimate of the maximum velocity the remnant material would approach if all the radioactive energy deposited in the remnant over the simulation time was converted into kinetic energy shared between all the remnant mass. The estimated velocities (between \tilda 500 to 800 km s$^{-1}$) are comparable to the velocity of a single grid cell for all models. However, it is likely that only the \nickel~rich part of the remnant envelope is ejected. The \cite{shen2017a} remnant model with most similar mass to our model remnants has an envelope mass of 0.1\,\msun~(see Section \ref{sec:remnant_implications} for more detailed discussion). If we assume all the radioactive energy is deposited in the remnant envelope during the simulation time, the envelope would reach an estimated velocity between \tilda 2000 to 2600 km s$^{-1}$ depending on the model. While more significant, these velocities are still only comparable to the velocities of a small number of inner grid cells in our simulations. As such, even if our model remnants become partially or fully unbound they will likely remain small on the scale of the computational grid. 

\subsection{Incorporating the remnant in \textsc{artis}}
\label{subsec:incorporating_the_remnant}
To calculate synthetic observables (light curves and spectra) for our models, to compare with observed SNe~Iax, we use the 3D time-dependent Monte Carlo radiative transfer code \artis\footnote{\href{https://github.com/artis-mcrt/artis/}{https://github.com/artis-mcrt/artis/}}~\citep{sim2007b, kromer2009a, bulla2015a, shingles2020a}. In the standard mode of operation, \artis~follows the propagation of $\gamma$-ray photons emitted by radioactive decays and deposits energy in the supernova ejecta before solving the radiative transfer problem self-consistently. For this study we also include the contribution from a luminous remnant source in our radiative transfer simulations by placing a \nickel~source in the centre of the models with \nickel~mass informed by the remnant \nickel~masses predicted by the L22 hydrodynamic explosion simulations. The main properties of the remnant and ejected material predicted by the L22 explosion simulations that form the basis of our new models are summarised in Table \ref{tab:model_properties}. As we move to fainter deflagration models the remnant \nickel~mass increases relative to the ejecta \nickel~mass. Therefore, if all energy released from the \nickel~decay chain in the remnant contributes to the optical display of the models the remnant will have a larger impact on the observed properties of faint models.

\begin{table}
    \centering
    \begin{tabular}{ccccc}
        \hline
        \rule[-1.5ex]{0pt}{0pt} 
        Model & $M_\mathrm{ej}$ & $M(\nickel)_\mathrm{ej}$ & $M_\mathrm{rem}$ & $M(\nickel)_\mathrm{rem}$ \\
        \hline
        r10\_d4.0\_Z & 0.227 & 0.092 & 1.16 & 0.054 \\
        r48\_d5.0\_Z & 0.054 & 0.018 & 1.34 & 0.038 \\
        r114\_d6.0\_Z & 0.014 & 0.0058 & 1.38 & 0.030 \\
        \hline
    \end{tabular}
    \caption{Summary of the total and \nickel~masses for the ejecta and remnant material of the L22 models which our new models are based on. Masses are given in solar masses.}
    \label{tab:model_properties}
\end{table}

\artis~is not configured to deal with a central luminous source in its standard mode of operation. Therefore, to incorporate a treatment for luminous remnants, we made use of developments made by \citet{collins2022a} which allows more flexibility in the way energy is injected in our \artis~simulations. We extended this method to treat energy due to radioactive decays from the \nickel~decay chain in the remnant separately to energy injected by radioactive decays in the ejecta. The rate and amount of radiation escaping the remnant is affected by a number of factors including the diffusion of the radiation through the remnant and potential mechanisms driven by the radioactive energy deposited in the remnant such as winds in the remnant envelope \citep{shen2017a} or the ejection of mass from the remnant \citep{maeda2022a}. However, for simplicity, in our prescription we assume the remnant emits instantaneously according to the \nickel~decay chain (i.e.\,assuming no significant delay due to diffusion in the remnant itself). Due to the compactness of the remnant material we expect the gamma rays emitted by the remnant to be trapped with their energy used to heat the remnant (see also discussion in K13). We therefore assume the remnant emits as a black body. Specifically, each time there is a \nickel~decay in the remnant, a packet of UVOIR radiation is emitted from the centre of the model grid and assigned a co-moving frame frequency sampled from a black body distribution with the temperature treated as a parameter of the simulation and representing the remnant temperature at the decay time (see discussion below). Once emitted, the UVOIR radiation packets travel through the ejecta and interact in the same way as UVOIR radiation packets originating in the ejecta. This approach allows us to obtain band light curves and spectra for our pure deflagration models with the contribution of a luminous remnant included, which we can compare directly to the light curves and spectra of observed SNe~Iax. 

In this work, we explore simulations first assuming a constant remnant temperature. We also test the sensitivity of our findings to the remnant temperature treatment adopted by exploring a temperature evolution of the remnant consistent with black body emission from a surface of constant radius. The SNe~Iax we compare to throughout this study have temperature estimates from observations which provide an approximate lower limit on the temperatures of potential remnants present for these objects. These temperature estimates guide the range of remnant temperatures we explore in this work: \cite{sahu2008a} and \cite{mccully2014b} estimate late time temperatures for the bright SNe~Iax, SN~2005hk of ${\lesssim} \, 4500\,\mathrm{K}$ and ${\sim} \, 4500$ to $7000\,\mathrm{K}$ respectively; \cite{maeda2022a} estimate a photospheric temperature for the intermediate-luminosity Type Iax, SN~2019muj of 5500\,K at 131 days after explosion and \cite{srivastav2022a} estimate the temperature of the faint SNe~Iax, SN~2020kyg  varies from ${\sim}\, 8000\,\mathrm{K}$ to $4000\,\mathrm{K}$ from \textit{g}-band peak to 60 days later. Additionally, models of post SNe~Iax remnants by \cite{shen2017a} with relatively similar remnant masses to those predicted for our models have estimated temperatures of ${\sim}\, 3000\,\mathrm{K}$ to $6000\,\mathrm{K}$ at 100 days after explosion.

Throughout this paper, the standard pure deflagration models retain their names as introduced by L22 with the structure rX\_dY\_Z where r refers to the ignition spark offset from the centre of the WD in km, d is the WD central density in $10^{9}$\,g\,$\mathrm{cm}^{-3}$ and Z the metallicity relative to the solar value (all models presented in this work adopt solar metallicity). We differentiate the models including a remnant contribution with an ‘‘R'' in the model name and also include information on the remnant temperature: for models with fixed remnant temperature we include this temperature in Kelvin in the model name and differentiate our model which adopts a temperature evolution consistent with a black body of constant radius with the ‘‘const\_rad'' label. Models which adopt remnant \nickel~masses less than those predicted by the L22 hydrodynamic explosion simulations of the models are labelled with the fraction of the predicted \nickel~mass that is adopted for the remnant. For example model r48\_d5.0\_Z\_R\_6000K\_0.33Ni is based on the standard r48\_d5.0\_Z pure deflagration model but includes the contribution from a remnant with a fixed characteristic temperature of 6000\,K and a \nickel~mass a third of that predicted by the L22 hydrodynamic explosion simulation of the standard deflagration model. We summarise the assumptions of the remnants for the models presented in this work in Table~\ref{tab:remnant_properties}.

\begin{table}
    \centering
    \begin{tabular}{ccc}
        \hline
        \rule[-1.5ex]{0pt}{0pt}
        Model & $M(\nickel)_\mathrm{rem}$ & $T_\mathrm{rem} $\,(K) \\
        \hline
        r10\_d4.0\_Z\_R\_2000K & 0.054 (1) & 2000 \\
        r10\_d4.0\_Z\_R\_8000K & 0.054 (1) & 8000 \\
        r10\_d4.0\_Z\_R\_15000K & 0.054 (1) & 15000 \\
        r10\_d4.0\_Z\_R\_const\_rad & 0.054 (1) & - \\
        r48\_d5.0\_Z\_R\_6000K\_0.33Ni & 0.013 ($\frac{1}{3}$) & 6000 \\
        r114\_d6.0\_Z\_R\_4000K & 0.030 (1) & 4000 \\
        r114\_d6.0\_Z\_R\_6000K\_0.1Ni & 0.003 ($\frac{1}{10}$) & 6000 \\
        \hline
    \end{tabular}
    \caption{Summary of the assumed remnant properties for our new simulations. The remnant \nickel~masses adopted are given in solar masses with the fraction of the remnant \nickel~mass predicted by L22 that these \nickel~masses represent included in brackets. The fixed remnant temperature is also stated for applicable models.}
    \label{tab:remnant_properties}
\end{table}

\subsection{Radiative transfer}
\label{subsec:RT_method}
\textsc{artis} follows the methods of \citet{lucy2002a, lucy2003a, lucy2005a} and is based on dividing the radiation field into indivisible energy packet Monte Carlo quanta. In this work we adopt the NLTE (non local thermodynamic equilibrium) approximation described by \citet{kromer2009a} and as used by F14 and L22. We do not make use of the full NLTE and non-thermal capabilities added to \artis~by \citet{shingles2020a} which are required to accurately model the nebular phase of SNe Ia (see, e.g., \citealt{dessart2014a, jerkstrand2017a, shingles2022a}). As such we do not extend the simulations presented in this work to nebular phases and halt our analysis at 80 days post explosion. We mapped the 3D abundance and density structure of the unbound material at the end of the L22 hydrodynamic simulations (by which point homologous expansion is a good approximation \citealt{roepke2005a}) to a $50^{3}$ Cartesian grid. 3D radiative transfer simulations were then carried out for each model. In each simulation, $3 \times 10^{7}$ energy packets were tracked through the ejecta for 200 logarithmically spaced time steps between 0.3 and 100 days. We used the atomic data set compiled by \citet{gall2012a} which is sourced from \citet{kurucz1995a} and \citet{kurucz2006a}. The much more extensive line list of \citet{kurucz2006a} is used for the important second and third ionisation stages of Fe, Co and Ni with the \citet{kurucz1995a} atomic data used for the remaining ions. We adopted a grey approximation in optically thick cells (c.f.\,\citealt{kromer2009a}) and at times earlier than 0.4 days post explosion, local thermodynamic equilibrium (LTE) was assumed. 

\section{Results}
\label{sec:results}
In the following section, we present comparisons between the bolometric and band light curves and spectra predicted by radiative transfer simulations of our new models with remnant contribution included and the standard L22 deflagration models along with observed SNe~Iax. To investigate the impact of including the remnant over a wide range of model brightnesses we include comparisons between bright, intermediate-luminosity and faint models along with corresponding observed SNe~Iax (the light curve properties of our new models and the L22 models we compare to are summarised in Tables \ref{tab:Bol_Sloan_lightcurve_properties} and \ref{tab:Bessel_lightcurve_properties}). The intermediate-luminosity and faint models with remnant contribution included that produced best agreement with observed SNe Iax adopt remnant \nickel~masses less than those predicted by the hydrodynamic explosion simulations of the models. We discuss this in detail in Section~\ref{sec:Discussion_Conclusions}. L22 found noticeable viewing angle effects for their standard pure deflagration models but these effects were not so large that they altered any of their findings. This is consistent with the viewing angle effects usually exhibited by pure deflagration models which are generally less substantial than those of other SNe Ia models due to their well-mixed ejecta structures (see e.g.\,K13, F14 and L22). We find our models including central remnants do not show significant viewing angle effects other than those already exhibited by the standard L22 pure deflagration models. Therefore, we focus on presenting angle-averaged results since these illustrate our main findings, but we comment on observer orientation effects where relevant. We also include the range in model parameters obtained from the line-of-site dependent light curves alongside the angle averaged values in Tables \ref{tab:Bol_Sloan_lightcurve_properties} and \ref{tab:Bessel_lightcurve_properties} for reference.  

\subsection{Comparisons with bright SNe~Iax}
\label{sec:Bright_Iax}

\subsubsection{Light curves}
\label{subsec:Bright_Iax_Lightcurves}
\begin{figure}
	\includegraphics[width=\columnwidth]{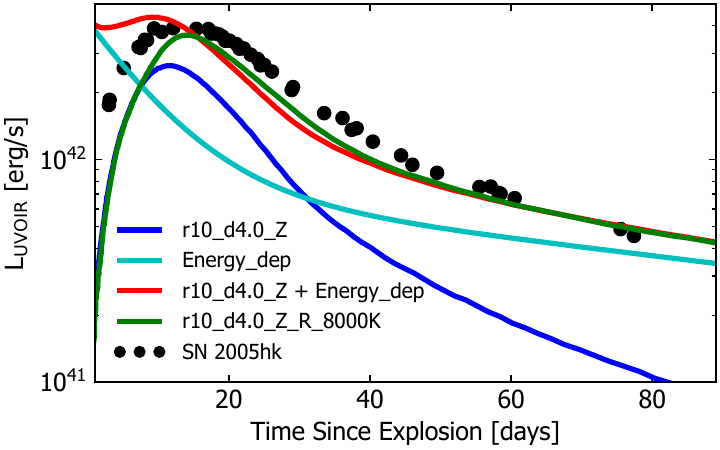}
    \caption{Angle averaged bolometric light curves for the r10\_d4.0\_Z L22 model and our new r10\_d4.0\_Z\_R\_8000K simulation. We also show the \nickel~decay energy deposition in the remnant (cyan), and the light curve obtained by adding this to the r10\_d4.0\_Z model light curve (red). The bolometric light curve of the bright SN~Iax, SN~2005hk \citep{phillips2007a} is shown for comparison.}
    \label{fig:BR_energy_dep}
\end{figure}

\begin{figure*}
	\includegraphics[width=\linewidth,trim={0.0cm 0.0cm 1.6cm 1.6cm},clip]{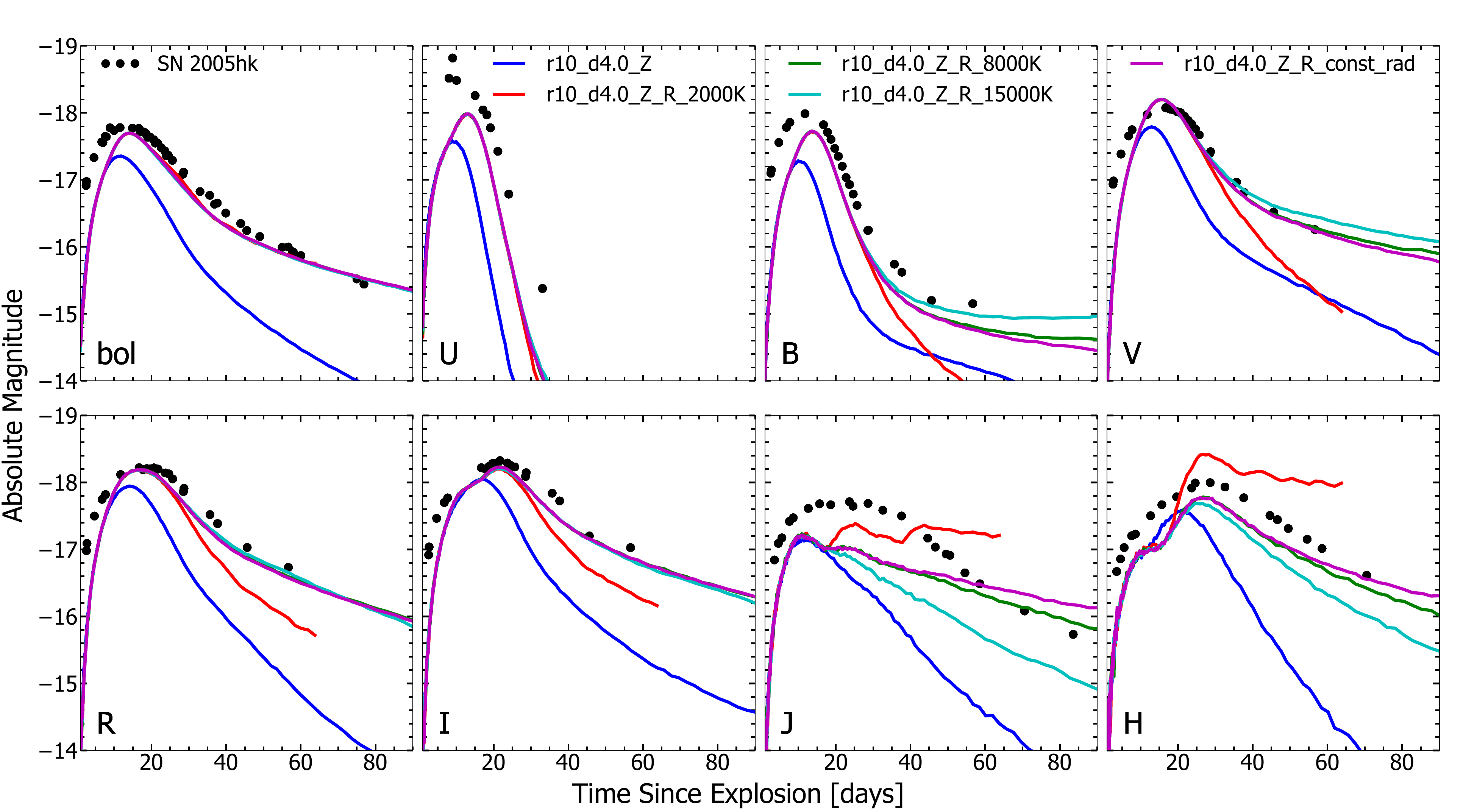}    
    \caption{Angle averaged bolometric and \textit{UBVRIJH}-band light curves for the r10\_d4.0\_Z L22 model and four new models from this work in which we also treat energy injected by radioactive decays in a luminous remnant in our radiative transfer simulations. Included for comparison are light curves of SN~2005hk \citep{phillips2007a}, a bright SN~Iax. As Model r10\_d4.0\_Z\_R\_2000K did not provide a well converged temperature solution at later times we only plot the light curves for this model up to 65 days. This poor convergence is a result of the low remnant temperature adopted for this model.}
    \label{fig:bright_band_lightcurves}
\end{figure*}

Figure~\ref{fig:BR_energy_dep} shows comparisons between the bolometric light curves of the brightest standard pure deflagration model from the L22 sequence, our new r10\_d4.0\_Z\_R\_8000K model and the bright SN~Iax, SN~2005hk. Also plotted is the bolometric light curve produced if energy deposited by radioactive decays in the remnant is instantaneously radiated (labelled ``Energy\_dep''). Following the approach of \cite{kromer2013a} we also plot the bolometric light curve produced if we simply add this remnant energy deposition to the bolometric light curve of the r10\_d4.0\_Z model (labelled \mbox{r10\_d4.0\_Z + Energy\_dep}). From Figure~\ref{fig:BR_energy_dep} we see the \mbox{r10\_d4.0\_Z + Energy\_dep} light curve produces good agreement with the bolometric light curve of SN~2005hk at later times (after ${\sim}\,50$ days). However, at early times it significantly over-estimates the brightness of the bolometric light curve until ${\sim}\,10$ days after explosion and then declines too quickly until ${\sim}\,40$ days. The r10\_d4.0\_Z\_R\_8000K bolometric light curve provides a good match to both the rise, decline and peak bolometric brightness of the SN~2005hk light curve, showing significantly improved agreement compared to the r10\_d4.0\_Z + R\_energy\_dep bolometric light curve at earlier times. Therefore, including the energy injected in the remnant directly in the radiative transfer simulations is important for accurate predictions of the remnant contribution to the early time bolometric light curves. Relative to Model r10\_d4.0\_Z the r10\_d4.0\_Z\_R\_8000K bolometric light curve provides a better match to SN~2005hk in rise and particularly in decline after peak also producing much better agreement with the peak brightness of SN~2005hk. We also note that the brightest viewing angles of Model r10\_d4.0\_Z\_R\_8000K reach the peak bolometric luminosity of the brightest members of the SNe~Iax class within their uncertainties, something not previously achieved by the L22 pure deflagration models that adopt a single-spark ignition.

Figure~\ref{fig:bright_band_lightcurves} shows angle-averaged bolometric and \textit{UBVRIJH}-band light curves for the r10\_d4.0\_Z model and four new models from this work in which energy injected by radioactive decays in the remnant are also included in the radiative transfer. The light curves of SN~2005hk are included for comparison. Each of the four models with remnant contribution included adopt the full remnant \nickel~mass predicted by the hydrodynamic explosion simulations of L22. However, the remnant temperature treatment is different for each model so we can investigate the sensitivity of the synthetic observables obtained from our radiative transfer simulations to what is assumed about the remnant's emission. The r10\_d4.0\_Z\_R\_2000K, r10\_d4.0\_Z\_R\_8000K and r10\_d4.0\_Z\_R\_15000K models all have fixed characteristic black body temperatures while the r10\_d4.0\_Z\_R\_const\_rad model corresponds to the remnant emitting as a black body surface of fixed radius (i.e.\,the remnant temperature decreases with time). We discuss how these different choices of remnant temperature treatment impact the band light curves of the models in detail below. However, we note that provided the energy is injected centrally in the remnant, the optical light curves are insensitive to different choices of remnant SED until \tilda 30 days after explosion (see Figure~\ref{fig:bright_band_lightcurves}). 

From Figure~\ref{fig:bright_band_lightcurves} we see that Model r10\_d4.0\_Z is not bright enough at peak and also too fast in both rise and decline in all bands to match the light curves of SN~2005hk. By comparison the r10\_d4.0\_Z\_R\_8000K, r10\_d4.0\_Z\_R\_15000K and r10\_d4.0\_Z\_R\_const\_rad models provide a significantly better match to the peak brightness, rise and decline of the band light curves of SN~2005hk (as was the case for the bolometric comparisons). Specifically, we emphasise the extremely good agreement between the \textit{R}, \textit{I} and \textit{H}-band model light curves and those of SN~2005hk, in particular in decline after peak. Compared to bright SNe~Iax such as SN~2005hk, previous pure deflagration models (e.g. the N5def model presented by K13) declined significantly too quickly after peak in these bands even when the decline was well matched in blue bands. Including the remnant contribution in our simulations significantly improves the agreement with the red optical and NIR bands although there is still a deficit in \textit{J}-band albeit much reduced.

The band light curves of the r10\_d4.0\_Z\_R\_8000K and r10\_d4.0\_Z\_R\_15000K models are very similar (see Figure \ref{fig:bright_band_lightcurves}). The biggest differences between the two models is in \textit{J} and \textit{H}-band where the r10\_d4.0\_Z\_R\_8000K model produces better agreement with SN~2005hk. This marginal improvement in agreement does not, however, rule out the higher remnant temperature of $15000\,\mathrm{K}$. Indeed, through additional simulations we have confirmed that a wide range of fixed remnant temperatures from $6000\,\mathrm{K}$ to $15000\,\mathrm{K}$ (the highest temperature we tested) produce model light curves in good agreement with those of SN~2005hk. Therefore, our results are not excessively impacted by different choices of fixed remnant temperature.  

Model r10\_d4.0\_Z\_R\_const\_rad allows us to test the impact of instead assuming the remnant radius is fixed, meaning the remnant temperature evolves in time. For Model r10\_d4.0\_Z\_R\_const\_rad we adopt a remnant radius of $4.9\times10^{14}\, \mathrm{cm}$. This is broadly consistent with the photospheric radius \cite{maeda2022a} derive for the proposed inner ejecta component of SN~2019muj, an intermediate luminosity SN~Iax. The remnant radius of Model r10\_d4.0\_Z\_R\_const\_rad yields a black body temperature of 8000\,K at 30 days after explosion (i.e.\,the time the optical bands become sensitive to the remnant temperature). We note that formally this radius is too large to be consistent with central energy injection in the remnant for early times. Nevertheless we continue to assume central energy injection noting the emerging optical light curves are insensitive to the assumed remnant SED prior to this time. Indeed, by the time the choice of remnant SED becomes relevant, it is already contained within the inner 15\% of the ejecta of Model r10\_d4.0\_Z\_R\_const\_rad which is consistent with our assumption of central energy injection in the remnant. This is true for all our models presented in this study that produce a reasonable match to observed SNe~Iax.

Comparing band light curves for the r10\_d4.0\_Z\_R\_const\_rad and r10\_d4.0\_Z\_R\_8000K models, we see adopting an evolving remnant temperature instead of a fixed temperature  leads to only very small variations in the band light curves, even at later times. The r10\_d4.0\_Z\_R\_const\_rad model shows a remnant effective temperature variation from ${\sim}\, 12500\, \mathrm{K}$ to $6000\,\mathrm{K}$ over the simulation time from 0.3 to 100 days. Our results are therefore not overly sensitive to remnant temperature variations in this range over the simulation time. The most noticeable differences are observed in \textit{J} and \textit{H}-bands where the r10\_d4.0\_Z\_R\_const\_rad model is slightly brighter than the r10\_d4.0\_Z\_R\_8000K model from \tilda 50 days onwards. This is because the r10\_d4.0\_Z\_R\_const\_rad model has lower effective remnant temperature at later times which increases the average wavelength of photon packets emitted by the remnant and thus the brightness in NIR bands. Overall, any variations in the band light curves between the r10\_d4.0\_Z\_R\_8000K and r10\_d4.0\_Z\_R\_const\_rad models are not significant in relation to comparisons made with the light curves of SN~2005hk even in \textit{J} and \textit{H}-bands. Therefore, within the time range of our simulations our conclusions are not sensitive to whether the remnant has a fixed or evolving temperature and we make the choice to primarily use remnants with fixed temperature for this study. 

Of the models with remnant contribution included Model r10\_d4.0\_Z\_R\_2000K produces by far the poorest agreement with SN~2005hk, declining much too quickly post peak in the optical bands and much too slowly after peak in the NIR bands. The poor agreement with the band light curves of SN~2005hk effectively rules out a remnant temperature of 2000\,K for our models. This remnant temperature being excluded is consistent with the ejecta temperatures estimated for SN~2005hk by \cite{sahu2008a} and \cite{mccully2014b} of $\lesssim$ 4500\,K and \tilda 4500 to 7000\,K respectively. These estimates do not, however, rule out the remnant temperatures adopted for the remaining models we compare to SN~2005hk in Figure~\ref{fig:bright_band_lightcurves}. 

\subsubsection{Spectra}
\label{subsec:Bright_Iax_Spectra}

\begin{figure*}
  \centering
  \includegraphics[width=.82\linewidth,trim={0.05 1.00cm 0 0},clip]{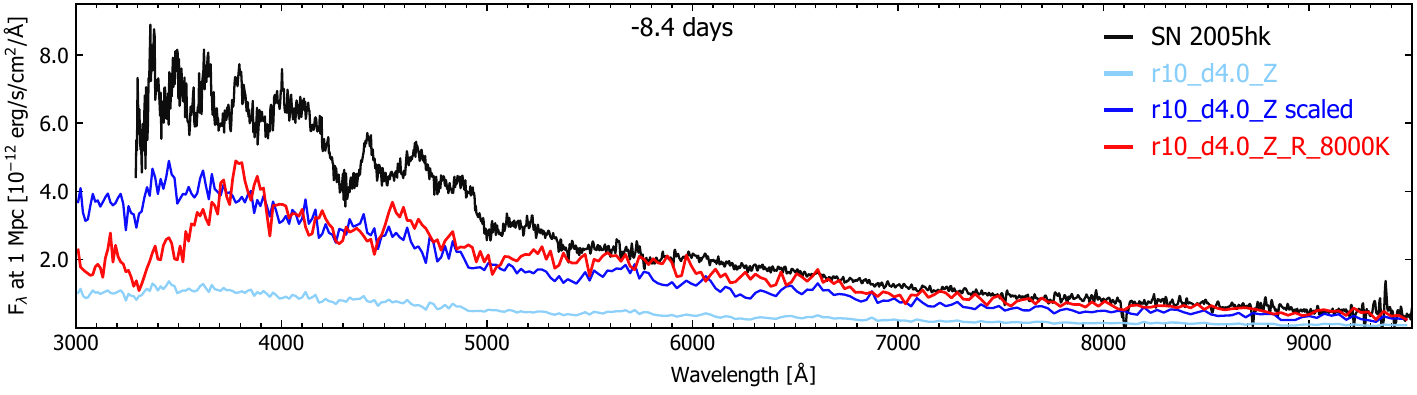}
  \includegraphics[width=.82\linewidth,trim={0.05 1.01cm 0 0},clip]{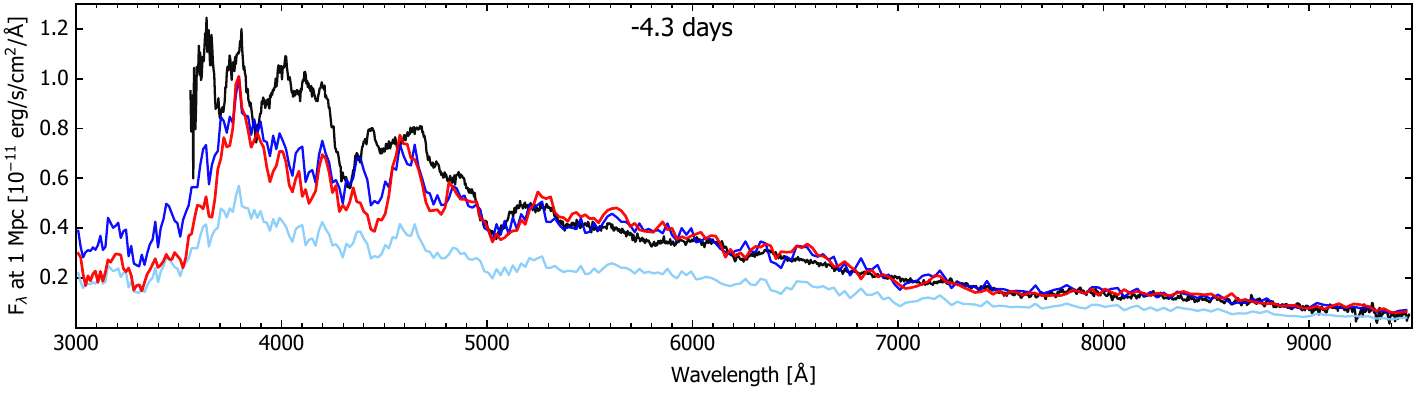}
  \includegraphics[width=.82\linewidth,trim={0.05 1.01cm 0 0},clip]{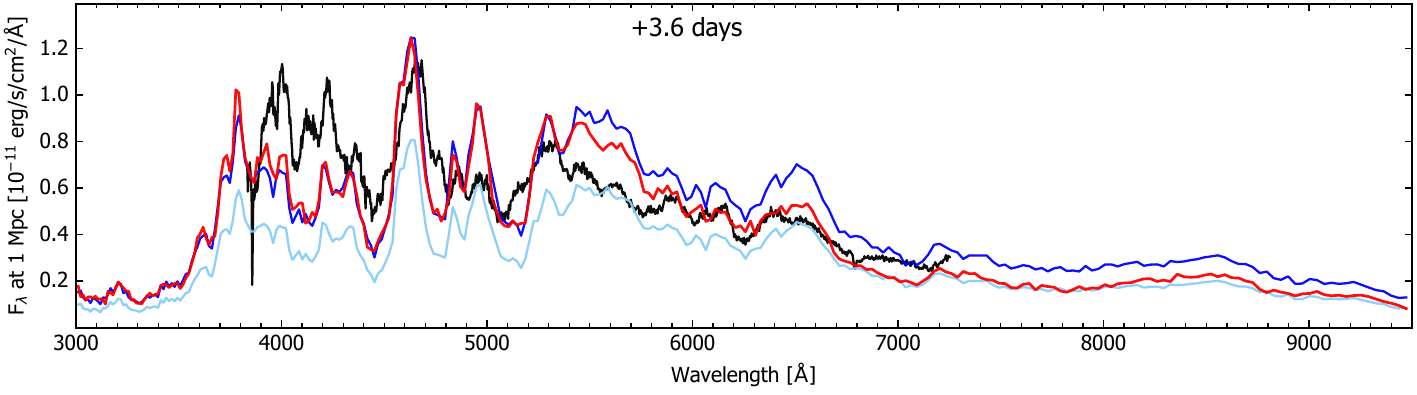}
  \includegraphics[width=.82\linewidth,trim={0.05 1.01cm 0 0},clip]{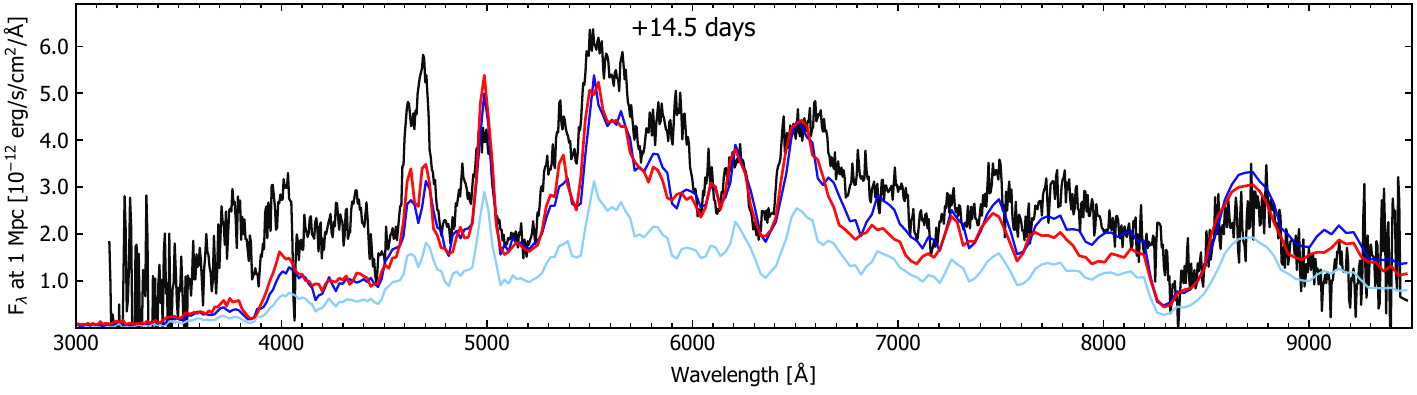}
  \includegraphics[width=.82\linewidth,trim={0.05 1.01cm 0 0},clip]{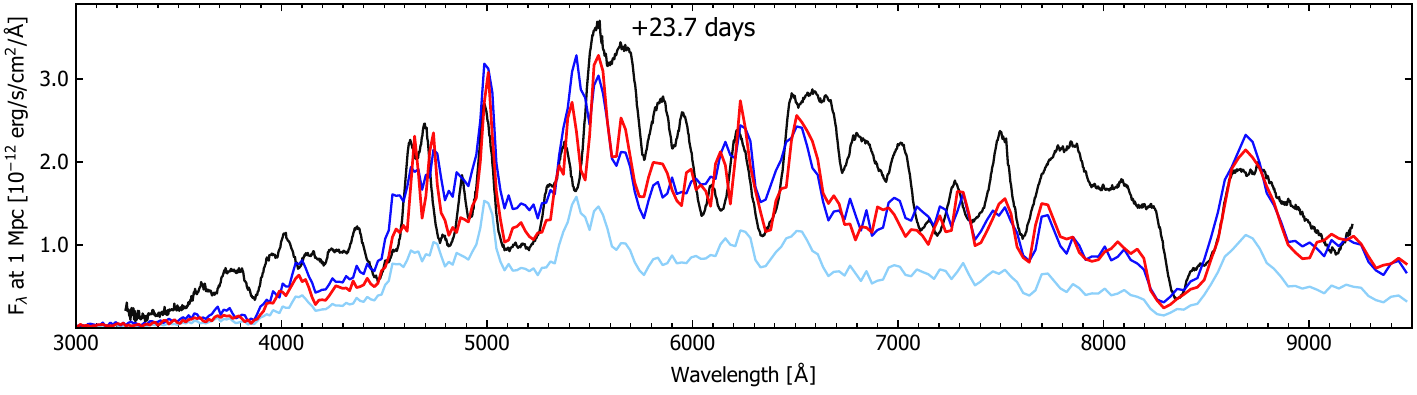}
  \includegraphics[width=.82\linewidth,trim={0.05 0.0cm 0 0},clip]{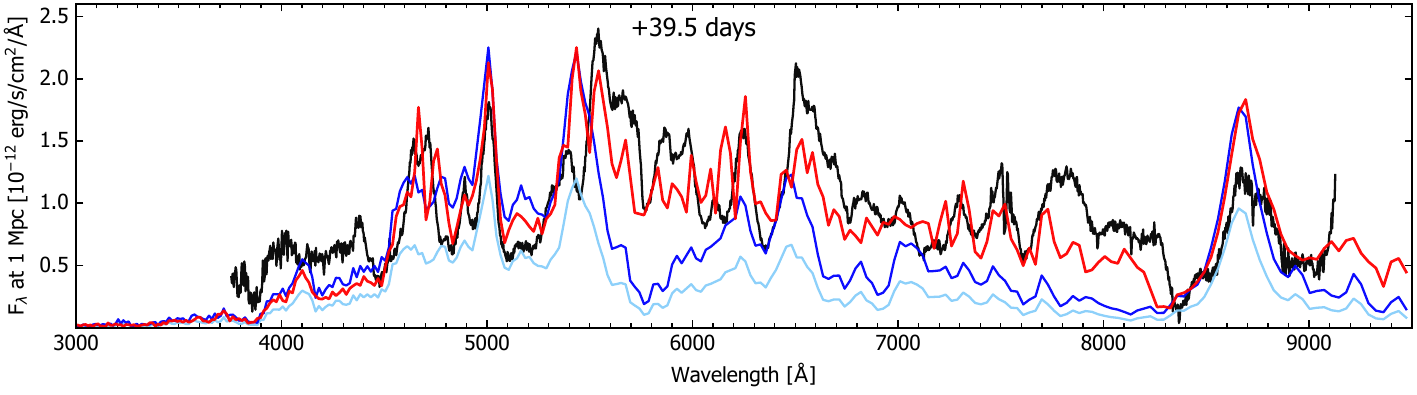}
  
  \caption{Absolute flux spectral comparisons between the angle averaged r10\_d4.0\_Z\_R\_8000K model and r10\_d4.0\_Z L22 model along with SN~2005hk \citep{phillips2007a}. Also plotted are the r10\_d4.0\_Z spectra scaled so their peak flux matches that of the r10\_d4.0\_Z\_R\_8000K model spectra. Times are relative to \textit{B}-band peak. The spectra of SN~2005hk have been corrected for reddening using the values from \citet{phillips2007a} and redshift corrected using the values from \citet{sahu2008a}.}
  \label{fig:bright_spectra}
\end{figure*}

Figure~\ref{fig:bright_spectra} shows spectral comparisons, at phases relative to \textit{B}-peak, in absolute flux, between Model r10\_d4.0\_Z\_R\_8000K, the r10\_d4.0\_Z L22 model and SN~2005hk. Also plotted are the r10\_d4.0\_Z spectra scaled by a constant factor across all wavelengths such that the peak of the r10\_d4.0\_Z spectra matches the peak of the r10\_d4.0\_Z\_R\_8000K model spectra at each epoch compared. All spectra scaled throughout this work follow this scaling approach.  We compare relative to \textit{B}-peak so the flux of the models is as similar as possible across the epochs compared. Compared to the r10\_d4.0\_Z model, the r10\_d4.0\_Z\_R\_8000K model spectra provide a significantly better match to the absolute flux of the SN~2005hk spectra across all epochs. This was already clear from the band light curve comparisons discussed above, but there are also some interesting differences in spectral shape and features, which we discuss below. 

The r10\_d4.0\_Z\_R\_8000K model shows its poorest agreement with the spectra of SN~2005hk when comparing at the earliest epoch 8.4 days before \textit{B}-peak. Although the SED of the r10\_d4.0\_Z\_R\_8000K model provides a reasonable match to SN~2005hk, particularly for wavelengths $\gtrsim5500\,$\r{A}, the spectrum is too faint. This is a result of the rapid early-time light curve evolution. When the spectra are instead compared at epochs relative to explosion (corresponding to a spectral epoch 1.3 days later than that plotted for Model r10\_d4.0\_Z\_R\_8000K) the flux match is significantly improved for this earliest epoch. From Figure~\ref{fig:bright_spectra} we also see the r10\_d4.0\_Z\_R\_8000K model spectrum shows a more noticeable lack of flux for wavelengths $\lesssim4000\,$\r{A}, with the r10\_d4.0\_Z scaled model producing better agreement with SN~2005hk over these wavelengths. This is a result of the shorter rise time to \textit{B}-peak of the r10\_d4.0\_Z model. This means the r10\_d4.0\_Z model is closer to explosion for the comparison shown and thus has hotter ejecta and therefore a bluer spectrum. When the models are instead compared relative to explosion they show extremely similar spectral evolution until \tilda 10 days since explosion. For this earliest epoch both models also show stronger spectral features for wavelengths $\lesssim5500\,$\r{A} when comparing relative to explosion, improving the spectral agreement with SN~2005hk. 

For the epochs at 4.3 days before \textit{B}-peak and 3.6 and 14.5 days after \textit{B}-peak the r10\_d4.0\_Z\_R\_8000K and r10\_d4.0\_Z scaled models display spectra which are very similar across all wavelengths. Both models display a reasonable match to the SED of SN~2005hk across these epochs. For the epochs at 4.3 days before and 3.6 days after \textit{B}-peak the agreement relative to the earliest epoch is primarily improved for wavelengths \mbox{$\lesssim5500\,$\r{A}}. However, at 14.5 days after \textit{B}-peak, the epoch the models show best overall spectroscopic agreement with SN~2005hk, there is an improved match to the SED of SN~2005hk across all wavelengths. Additionally, over these epochs, both models show improved agreement with the spectra of SN~2005hk relative to the earliest epoch shown due to the increased number and strength of spectral features.

At 23.7 days after \textit{B}-peak, the model spectra remain very similar although the r10\_d4.0\_Z\_R\_8000K model matches the strengths of some of the absorption features in the blue wavelengths observed for SN~2005hk better, such as the strong absorption trough at $\sim5200\,$\r{A} primarily due to Fe \textsc{ii}. While the overall spectral agreement with SN~2005hk is still good for this epoch, both models show emission features (primarily attributed to Fe and Co \textsc{ii} in the models) that are too weak relative to SN~2005hk, particularly for red wavelengths. For the earlier epochs shown, the model spectral features are in general more blue-shifted than those of SN~2005hk. However, for later epochs, although there are still some mismatches, the model spectral features do not show a preference for being blue-shifted or red-shifted relative to SN~2005hk. This may suggest the outer regions of the model ejecta, where spectra are formed at early times, have velocities in general too high compared to SN~2005hk, while the inner ejecta of the models, have velocities which provide a better match to SN~2005hk. Another explanation for this behaviour may be differences in the ionisation state between the outer ejecta of SN~2005hk and the models leading to the early-time spectral features of the models generally being formed at higher velocities. 

The largest difference between the r10\_d4.0\_Z\_R\_8000K and r10\_d4.0\_Z scaled model spectra is found for the latest epoch shown (39.5 days after \textit{B}-peak). At this epoch, the r10\_d4.0\_Z\_R\_8000K spectrum matches the SED of SN~2005hk well across all wavelengths shown and reproduces the strength and velocities of many of the spectral features across a wide range of wavelengths. The r10\_d4.0\_Z scaled spectrum, on the other hand, has a noticeable lack of flux for wavelengths longer than $\sim5500\,$\r{A} and has spectral features much weaker than SN~2005hk for these redder wavelengths. This lack of flux for red wavelengths at later times is a general issue for standard pure deflagration models (see Section \ref{sec:intro}). Overall, the r10\_d4.0\_Z\_R\_8000K and r10\_d4.0\_Z scaled model spectra are very similar for all but the latest epoch shown, producing similarly good spectroscopic agreement with SN~2005hk over these epochs. 

\subsection{Comparisons with intermediate-luminosity SNe~Iax}
\label{sec:Intermediate_Iax}
\subsubsection{Light curves}
\label{subsec:Intermediate_Iax_Lightcurves}
\begin{figure*}
	\includegraphics[width=\linewidth,trim={0.4cm 0.1cm 1.8cm 1.8cm},clip]{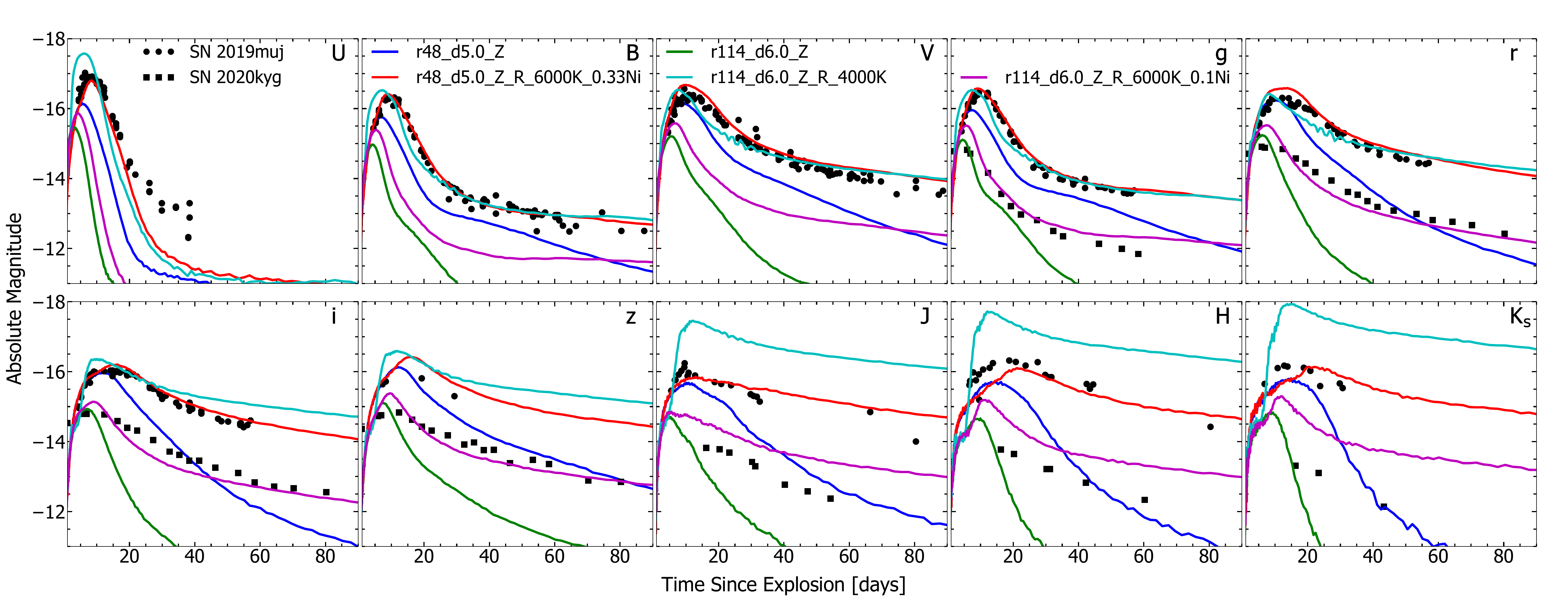}    
    \caption{Angle averaged \textit{UBVJHK}\textsubscript{s} and \textit{griz} band light curves for the r48\_d5.0\_Z\_R\_6000K\_0.33Ni, r114\_d6.0\_Z\_R\_4000K and r114\_d6.0\_Z\_R\_6000K\_0.1Ni models along with the L22 r48\_d5.0\_Z and r114\_d6.0\_Z models. The light curves of SN~2019muj \citep{barna2021a, kawabata2021a}, an intermediate-luminosity SN~Iax and the faint SN~Iax, SN~2020kyg \citep{srivastav2022a} are included for comparison. Note, that the \textit{JHK}\textsubscript{s} magnitudes are from the 2MASS catalogue (\citealt{persson1998a}). The \textit{K}\textsubscript{s} filter is a \textit{K} filter which has been modified to reduce thermal background for ground based telescopes.} 
    \label{fig:intermediate_faint_band_lightcurves}
\end{figure*}

Figure~\ref{fig:intermediate_faint_band_lightcurves} shows band light curves plotted for Model r48\_d5.0\_Z\_R\_6000K\_0.33Ni (which adopts a remnant \nickel~mass a third of that predicted by L22, see Table \ref{tab:remnant_properties} for summary of remnant properties), Model r114\_d6.0\_Z\_R\_4000K, the L22 r48\_d5.0\_Z model and the intermediate-luminosity SN~Iax, SN~2019muj. Model r48\_d5.0\_Z\_R\_6000K\_0.33Ni, our model with remnant contribution included in best agreement with SN~2019muj, produces significantly better agreement with the light curves of SN~2019muj than the closest matching L22 model r48\_d5.0\_Z, matching the peak, rise and particularly decline of SN~2019muj better in all bands (apart from $z$-band, where the small number of photometric points for SN~2019muj makes the level of agreement hard to judge between models). In particular, the remnant contribution removes the systematic problem of the light curves declining too rapidly. The improvement is again most dramatic in the red optical and NIR bands. While the r48\_d5.0\_Z\_R\_6000K\_0.33Ni model produces best agreement with the optical band light curves of SN~2019muj, the agreement is still reasonably good for all NIR bands shown, despite the rise being a bit slow relative to SN~2019muj in the \textit{H} and \textit{K}\textsubscript{s}-bands. The remnant temperature adopted for Model r48\_d5.0\_Z\_R\_6000K\_0.33Ni is consistent with the photospheric temperature of \tilda 5500 K estimated by \cite{maeda2022a} for SN~2019muj at 131 days after explosion. Overall, including energy injected by radioactive decays in the remnant significantly improves the agreement of pure deflagration models with the band light curves of intermediate-luminosity SNe~Iax such SN~2019muj. 

To match the optical luminosity of SN~2019muj for models including the entire remnant \nickel~mass predicted by L22 we require remnant temperatures lower than that adopted for Model r48\_d5.0\_Z\_R\_6000K\_0.33Ni. Of the models we investigated that adopted the entire L22 \nickel~mass, Model r114\_d6.0\_Z\_R\_4000K produces best agreement with SN~2019muj. Comparing the light curves of this model and Model r48\_d5.0\_Z\_R\_6000K\_0.33Ni to SN~2019muj we see both models produce good agreement with the \textit{BVgr} band light curves of SN~2019muj although Model r48\_d5.0\_Z\_6000K\_0.33Ni provides an overall better match in these bands. Model r48\_d5.0\_Z\_6000K\_0.33Ni also provides a reasonably good match to SN~2019muj in all other bands shown. However, Model r114\_d6.0\_Z\_R\_4000K is too bright at peak in all remaining bands relative to SN~2019muj. Additionally, all the NIR light curves of Model r114\_d6.0\_Z\_R\_4000K are at least a magnitude too bright compared to SN~2019muj from peak until the end of the simulation at 100 days. Model r114\_d6.0\_Z is the faintest from the L22 sequence meaning it is not possible to reproduce the correct brightness for any observed SNe~Iax fainter than SN~2019muj if we assume the entire remnant \nickel~mass contributes directly to the observed luminosity of the model. Therefore, to produce good agreement with both intermediate-luminosity and faint SNe~Iax, we must adopt remnant \nickel~masses in our models less than those predicted by L22. We discuss the implications of this finding in more detail in Section \ref{sec:Discussion_Conclusions}.  

\subsubsection{Spectra}
\label{subsec:Intermediate_Iax_Spectra}
\begin{figure*}
  \centering
  \includegraphics[width=.82\linewidth,trim={0.05 1.00cm 0 0},clip]{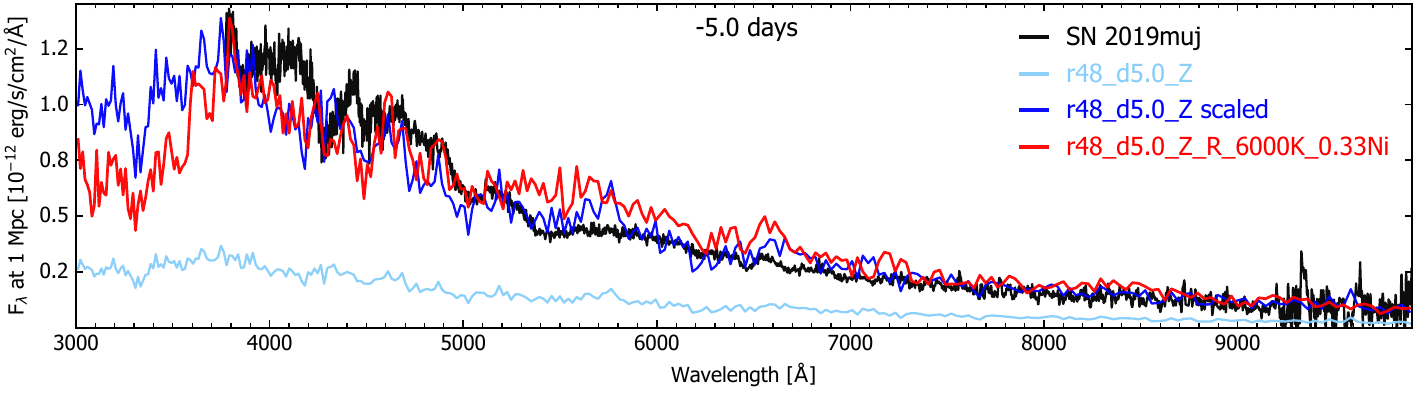}
  \includegraphics[width=.82\linewidth,trim={0.05 1.01cm 0 0},clip]{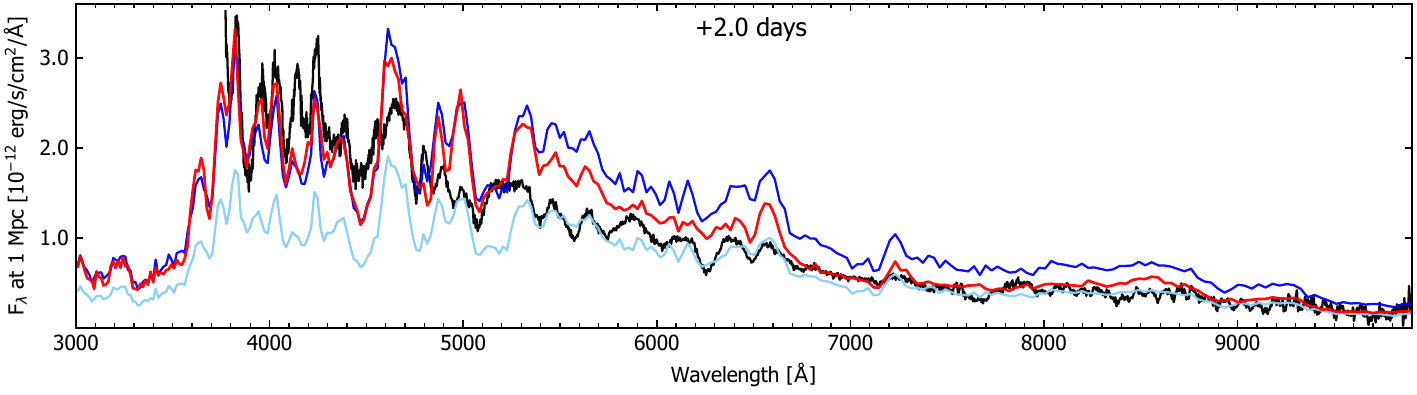}
  \includegraphics[width=.82\linewidth,trim={0.05 1.01cm 0 0},clip]{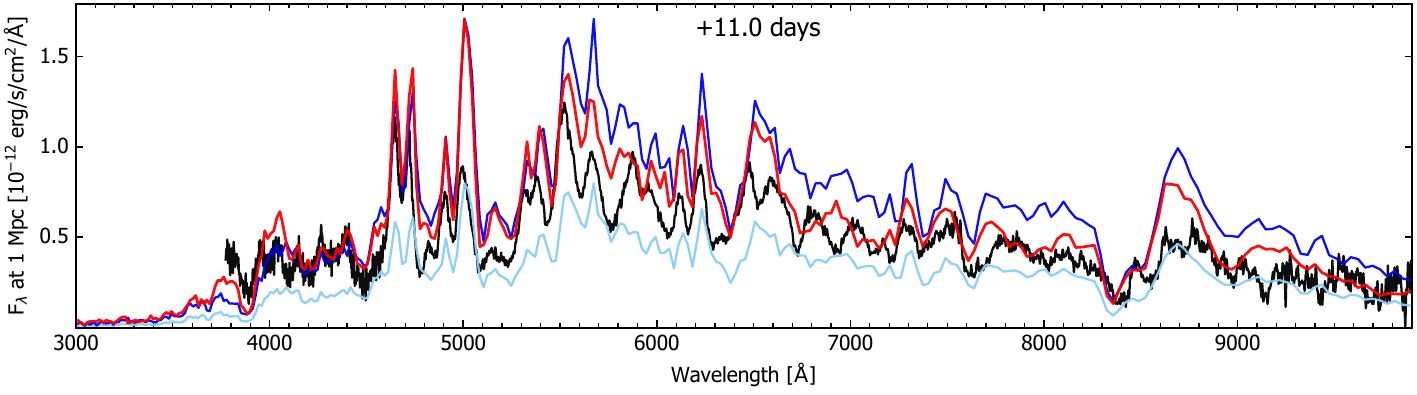}
  \includegraphics[width=.82\linewidth,trim={0.05 1.01cm 0 0},clip]{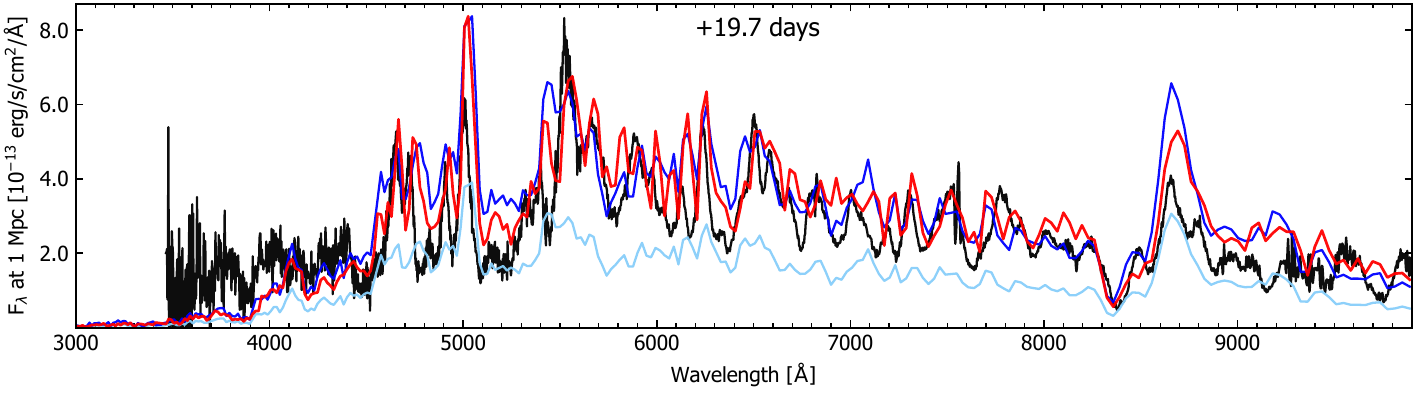}
  \includegraphics[width=.82\linewidth,trim={0.05 1.01cm 0 0},clip]{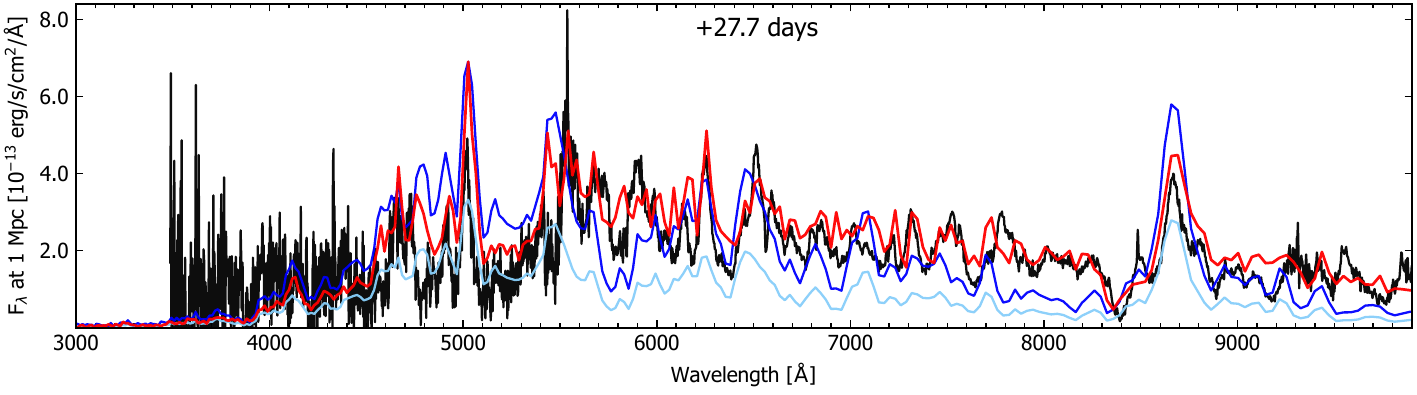}
  \includegraphics[width=.82\linewidth,trim={0.05 0.0cm 0 0},clip]{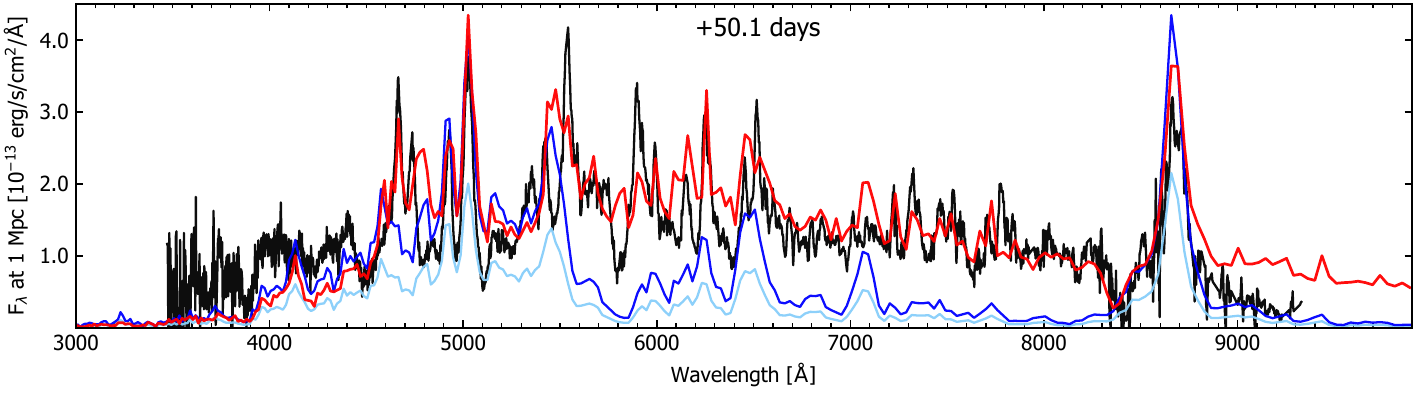}
  
  \caption{Absolute flux spectral comparisons between the angle averaged r48\_d5.0\_Z\_6000K\_0.33Ni model, r48\_d5.0\_Z L22 model and SN~2019muj \citep{barna2021a}. Also plotted are the r48\_d5.0\_Z model spectra scaled so their peak flux matches that of Model r48\_d5.0\_Z\_6000K\_0.33Ni. Times are relative to \textit{B}-band peak. The spectra of SN~2019muj have been corrected for red shift and reddening using the values from \citet{barna2021a}. The observed spectra shown here were taken from WISeREP (\citealt{yaron2012a}).}
  \label{fig:intermediate_spectra}
\end{figure*}

Figure~\ref{fig:intermediate_spectra} shows absolute flux spectral comparisons between Model r48\_d5.0\_Z\_6000K\_0.33Ni, Model r48\_d5.0\_Z, Model r48\_d5.0\_Z scaled to match the peak of r48\_d5.0\_Z\_6000K\_0.33Ni and SN~2019muj. Model r48\_d5.0\_Z\_6000K\_0.33Ni provides a significantly better overall flux match to the spectra of SN~2019muj at all epochs compared to the r48\_d5.0\_Z model. Therefore, following the same approach as Section \ref{subsec:Bright_Iax_Spectra} we primarily focus on the r48\_d5.0\_Z scaled flux model alongside Model r48\_d5.0\_Z\_6000K\_0.33Ni for our comparisons to SN~2019muj so we can evaluate how including the remnant contribution impacts the spectral evolution of intermediate-luminosity deflagration models. 

The r48\_d5.0\_Z\_6000K\_0.33Ni model provides a very good match to the overall SED and the strengths and velocities of a significant number of the spectral features of SN~2019muj, across all epochs shown. Relative to the r48\_d5.0\_Z\_6000K\_0.33Ni model, the r48\_d5.0\_Z scaled model produces a similarly good match to the spectral features of SN~2019muj  except for the latest epoch shown, where the r48\_d5.0\_Z spectral features appear too weak for wavelengths longer than $\sim6500\,$\r{A} primarily due to the lack of flux for the model at these red wavelengths. The r48\_d5.0\_Z scaled model produces a significantly worse match to the colour evolution of SN~2019muj: while the model matches the first epoch shown (5 days before \textit{B}-peak), it becomes too red at 2 and 11 days after peak. At 19.7 days after peak the model then exhibits an improved match to the colours of SN~2019muj before showing spectra which become increasingly too blue for the latest two epochs shown at 27.7 and 50.1 days after peak. Compared to bright models, including the remnant contribution for intermediate-luminosity models leads to a greater change in spectral properties and bigger overall improvement in the agreement with observed SNe~Iax. This suggests the energy injected into the ejecta due to the radiation emitted from the remnant has a greater impact on the ejecta conditions of intermediate-luminosity deflagration models.

\subsection{Comparisons with faint SNe~Iax}
\label{sec:Faint_Iax}
\subsubsection{Light curves}
\label{subsec:Faint_Iax_Lightcurves}
From Figure~\ref{fig:intermediate_faint_band_lightcurves} we can see band light curve comparisons between Model r114\_d6.0\_Z\_6000K\_0.1Ni (which adopts a remnant \nickel~mass a tenth of that predicted by L22), the r114\_d6.0\_Z L22 model and the faint SNe~Iax, SN~2020kyg. From fitting the SED of SN~2020kyg, \cite{srivastav2022a} find a black body temperature which varies from \tilda 8000 K at g band peak to \tilda 4000 K at 60 days after g band peak, which is consistent with the remnant temperature we adopt for our model. L22 predict a remnant \nickel~mass \tilda 5 times greater than the ejecta \nickel~mass for Model r114\_d6.0\_Z. However,  the ejecta component of Model r114\_d6.0\_Z is already brighter than SN~2020kyg at peak in most bands. Therefore, it follows that to provide a reasonable match to the peak brightness of SN~2020kyg our model with remnant contribution included must adopt a \nickel~mass only a small fraction of that predicted by L22 (see Section \ref{sec:Discussion_Conclusions} for discussion).

The r114\_d6.0\_Z\_6000K\_0.1Ni model matches the decline of SN~2020kyg in all bands better than the standard r114\_d6.0\_Z model, which  declines much too rapidly in all bands. In particular the r114\_d6.0\_Z\_6000K\_0.1Ni model quite successfully matches the decline after peak in the optical bands until \tilda 80 days after explosion (the latest photometric point of SN~2020kyg available). However, the r114\_d6.0\_Z\_6000K\_0.1Ni model is too bright at peak by up to half a magnitude in the optical bands and a magnitude or more in the NIR bands, while also declining too slowly after peak in the NIR bands. Although we expected the r114\_d6.0\_Z\_6000K\_0.1Ni model would be too bright compared to SN~2020kyg, the more significant overproduction of flux at NIR wavelengths suggests our adopted remnant temperature may be too low. 

To test this we carried out another simulation also based on Model r114\_d6.0\_Z but with an increased fixed remnant black body temperature of $8000\,\mathrm{K}$ and a further reduced \nickel~mass a twentieth of that predicted by L22. This model did produce light curves which were slightly fainter at peak, up to \tilda 0.2 magnitudes in optical bands and as much as \tilda 0.4 in NIR bands, leading to closer agreement with SN~2020kyg. This model also produced NIR band light curves that matched the decline SN~2020kyg more successfully. However, this came at the expense of significantly reducing the agreement of the model decline in the \textit{r}, \textit{i} and \textit{z}-bands and in general providing a worse match to the band light curves of SN~2020kyg compared to the r114\_d6.0\_Z\_6000K\_0.1Ni model. We also note that as we move to faint deflagration models in which we include the remnant contribution, the models become more sensitive to the choice of remnant temperature. This may be a result of the ejecta becoming optically thin more rapidly for faint models meaning the remnant radiation has a larger contribution to the synthetic observables at earlier times for faint models.

Overall, we have been able to achieve a significant improvement in agreement with the band light curves of SN~2020kyg, particularly in decline, by including the remnant radiation in our simulations. However, there is no obvious way using our approach to produce a simulation based on the standard L22 pure deflagration models with remnant contribution included that can produce good agreement with the light curves of SN~2020kyg across all bands (see Section \ref{sec:Discussion_Conclusions} for further discussion).

\subsubsection{Spectra}
\label{subsec:Faint_Iax_Spectra}
\begin{figure*}
  \centering
  \includegraphics[width=.82\linewidth,trim={0.05 1.00cm 0 0},clip]{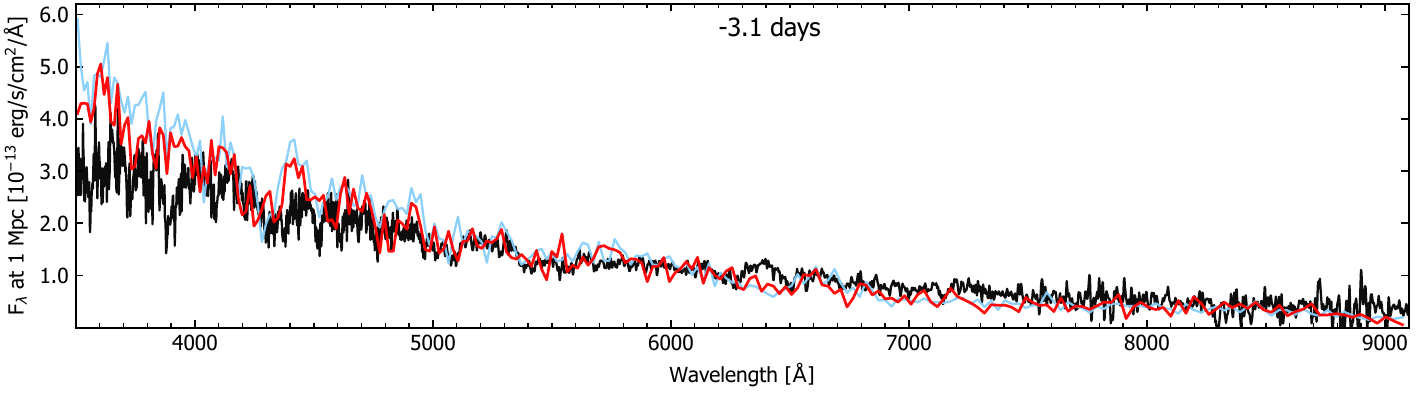}
  \includegraphics[width=.82\linewidth,trim={0.05 1.01cm 0 0},clip]{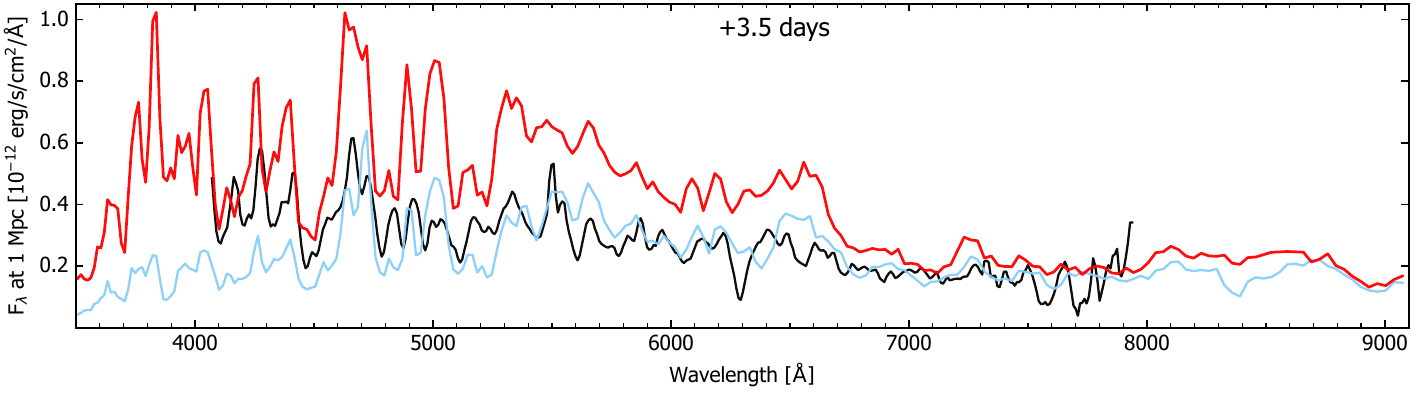}
  \includegraphics[width=.82\linewidth,trim={0.05 1.01cm 0 3},clip]{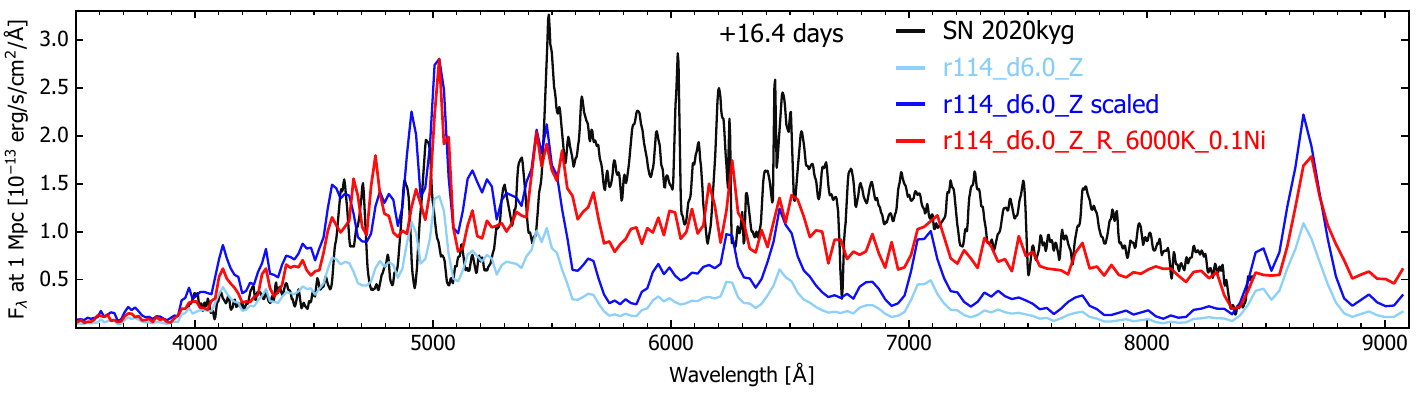}
  \includegraphics[width=.82\linewidth,trim={0.05 1.01cm 0 0},clip]{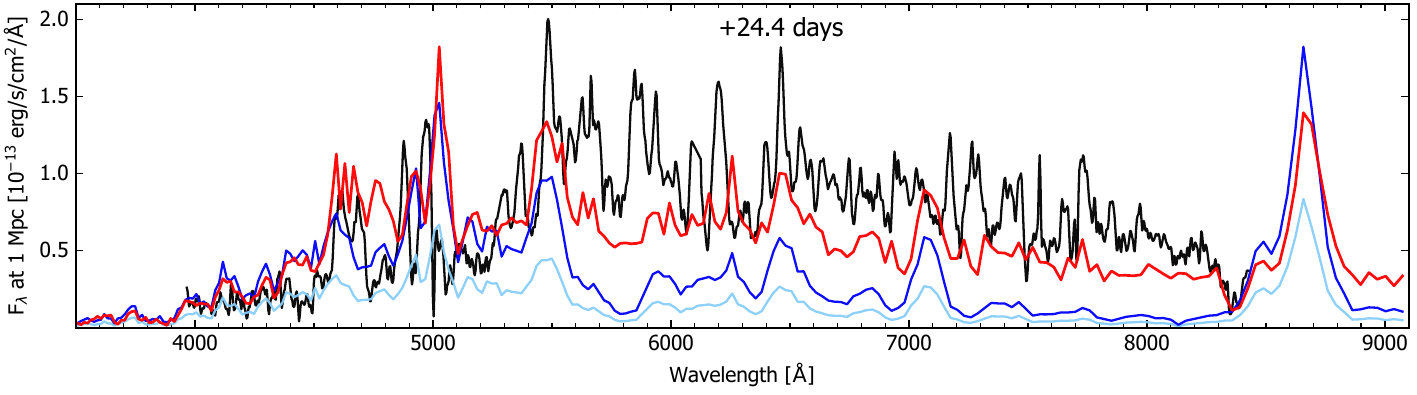}
  \includegraphics[width=.82\linewidth,trim={0.05 0.0cm 0 0},clip]{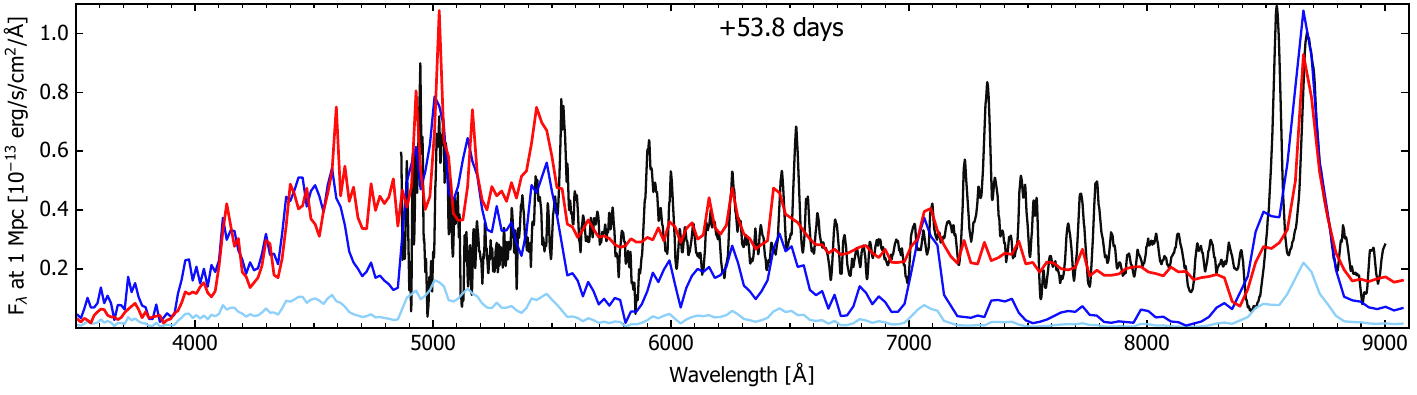}
  \caption{Absolute flux spectral comparisons between the angle averaged r114\_d6.0\_Z\_6000K\_0.1Ni model, r114\_d6.0\_Z L22 model and SN~2020kyg \citep{srivastav2022a}. Also plotted are the r114\_d6.0\_Z spectra scaled so their peak flux matches that of Model r114\_d6.0\_Z\_6000K\_0.1Ni where required, to allow clearer comparison of  spectral features. Times are relative to \textit{g}-band peak. The spectra of SN~2020kyg have been corrected for red shift and reddening using the values from \citet{srivastav2022a}.}
  \label{fig:faint_spectra}
\end{figure*}

Figure~\ref{fig:faint_spectra} shows spectroscopic comparisons between Model r114\_d6.0\_Z\_6000K\_0.1Ni, Model r114\_d6.0\_Z, Model r114\_d6.0\_Z scaled to match Model r114\_d6.0\_Z\_6000K\_0.1Ni at peak and SN~2020kyg. For the earliest epoch shown (3.1 days before g peak) both models are spectroscopically very similar and provide a good match in absolute flux to the spectrum of SN~2020kyg. At the next epoch shown (3.5 days after peak) the r114\_d6.0\_Z model still shows a good match to the absolute flux of SN~2020kyg while the r114\_d6.0\_Z\_6000K\_0.1Ni model becomes too bright. For the last 3 epochs shown (16.4, 24.4 and 53.8 days after g peak) the r114\_d6.0\_Z\_6000K\_0.1Ni model provides a significantly better match to the absolute flux of SN~2020kyg compared to the r114\_d6.0\_Z model which becomes much too faint. Additionally, while the spectra of the r114\_d6.0\_Z\_6000K\_0.1Ni model are also slightly too blue they provide a significantly better match to the overall colours: the r114\_d6.0\_Z model is significantly too blue. Therefore, as was the case for intermediate-luminosity deflagration models, including the remnant radiation for faint deflagration models improves agreement with observed SNe~Iax because of a wavelength dependent increase in flux for the r114\_d6.0\_Z\_6000K\_0.1Ni model, producing a colour evolution more similar to SN~2020kyg. 

The r114\_d6.0\_Z\_6000K\_0.1Ni model is able to reproduce the strength and location of many of the spectral features of SN~2020kyg for the earlier epochs shown. However, for the latest two epochs shown, while the model is still reproducing some of the spectral features, they are often too weak. This is particularly clear for the last epoch shown at wavelengths redder than $\sim5500\,$\r{A} where the r114\_d6.0\_Z\_6000K\_0.1Ni model spectra struggles to match the large number of strong sharp features exhibited by the spectrum of SN~2020kyg. This is likely because the spectra of SN~2020kyg at these later epochs show an increasing number of features due to forbidden transitions \citep{srivastav2022a} which can not be treated accurately using the atomic data set and approximate NLTE treatment we have adopted in our radiative transfer simulations for this work. Adopting the full NLTE treatment and more extensive atomic dataset of \cite{shingles2020a} may produce model spectra which more successfully reproduce these spectral features at later times. 

Focusing again on the latest two epochs shown we see interesting differences in the evolution of the Ca \textsc{ii} NIR triplet. In the spectral comparison at 24.4 days after peak the absorption feature at $\sim8400\,$\r{A}, which we have confirmed to be the most blue-ward absorption feature of the Ca \textsc{ii} NIR triplet in the models, matches the velocity of the feature in SN~2020kyg well. However, for the spectral comparison at 53.8 days after peak while this absorption feature has remained at the same velocity for the model it has shifted significantly to the red for SN~2020kyg.  Additionally, comparing the overall profile of the Ca \textsc{ii} NIR triplet in the latest epoch shown we see that it is both bluer and broader for the deflagration models compared to SN~2020kyg. This explains the lack of emission from the 8542\,\r{A} component of the Ca \textsc{ii} NIR triplet for the models as it is suppressed by the blue shifted absorption tail of the 8662\,\r{A} component, something which \cite{maeda2022a} also observed for their spectral models of SN~2019muj. These comparisons of the Ca \textsc{ii} NIR triplet suggest there is a low velocity, higher density region of the ejecta in SN~2020kyg, where spectral features are forming at later epochs, which is not accounted for in our models. Such an ejecta structure has already been proposed for other observed SNe~Iax (see e.g. \citealt{sahu2008a}) while \cite{foley2016a} and \cite{shen2017a} suggest such an ejecta structure may be consistent with radioactively driven post SNe~Iax remnant winds. Alternatively, \cite{maeda2022a} have suggested that a scenario involving a secondary higher density ejecta component which is ejected from the remnant envelope \tilda a month after explosion may explain this spectral behaviour at later times. 

\section{Discussion and Conclusions}
\label{sec:Discussion_Conclusions}
\subsection{Summary}
\label{sec:Summary}
We have presented a new set of radiative transfer simulations for models based on the \mch~CO WD pure deflagration models of L22 in which we also include energy injection from a luminous remnant, as predicted in the explosion models and possibly observed for the SNe~Iax, SN~2008ha \citep{foley2014a}, SN~2012Z \citep{mccully2022a}, SN~2014dt \citep{kawabata2018a} and SN~2019muj \citep{kawabata2021a}. While our models require an assumption about the SED of photons from the remnant we found no significant differences in the synthetic observables of the models where we assumed a black body SED with either constant radius or temperature. Therefore, our models are not overly sensitive to the specific choice of effective remnant temperature at a given time. Moreover, our models which produced best agreement with observed SNe~Iax all adopted remnant temperatures consistent with late time temperature estimates of observed SNe~Iax. We note, however, that if we were to extend our simulations to later times, the remnant temperature predicted for a model with fixed remnant radius will increasingly diverge from a fixed remnant temperature making this choice more relevant. Additionally, at later times the ejecta become significantly more optically thin meaning radiation emitted from the remnant will be less impacted by interaction with ejecta material and thus will likely become more sensitive to the remnant temperature treatment utilised.

Including the remnant \nickel~mass predicted for the brightest pure deflagration model from the L22 sequence increases the peak bolometric magnitude of the model by \tilda half a magnitude. This means the brightest viewing angles of this model can match the peak bolometric luminosity of the brightest SNe~Iax within their uncertainties, something not previously possible for the L22 single-spark pure deflagration models. Additionally, the light curves are in good agreement with the bright SNe~Iax, SN~2005hk in peak magnitude, rise and decline in all bands. This represents a significant improvement as previous bright deflagration models presented by K13, K14 and L22 exhibited declines systematically too rapid compared to bright SNe~Iax in the red optical and NIR bands. The inclusion of the remnant also improved spectroscopic agreement with SN~2005hk, primarily due to the better overall flux agreement, although for the latest epoch compared the redder spectrum formed also improved the match with the spectral colours of SN~2005hk. However, including the remnant contribution causes a more pronounced change in ejecta conditions for intermediate-luminosity and faint models leading to bigger improvements in their spectroscopic agreement with observed SNe~Iax relative to standard deflagration models.

Of the models investigated in this work, intermediate-luminosity deflagration models with remnant contribution included display the largest improvement in agreement with observed SNe~Iax relative to standard deflagration models. Including the remnant produces model light curves in good agreement with the intermediate-luminosity SN~Iax, SN~2019muj in terms of peak magnitude, rise and decline in all bands. This again represents a significant improvement compared to standard deflagration models, which show light curves that decline systematically too fast after peak in all bands relative to observed SNe~Iax. Additionally, including the remnant contribution leads to significantly improved spectral agreement with SN~2019muj due to the better match with the evolution of the SED. 

Including the remnant contribution also significantly improves the agreement with the light curve evolution of the faint SNe~Iax, SN~2020kyg. This improvement is primarily due to the slower light curve decline after peak in all bands compared to faint standard deflagration models which show declines that are significantly too fast relative to SN~2020kyg. Additionally, as was the case for intermediate-luminosity models, including the remnant contribution leads to significantly better spectral agreement with SN~2020kyg due to the improved match to the evolution of its SED. The inclusion of the remnant does, however, make the model light curves too bright at peak relative to SN~2020kyg in all bands and significantly too bright over the whole simulation time in NIR bands. This is to be expected as the faintest member of the L22 model sequence is already slightly brighter than SN~2020kyg at peak. Therefore, to achieve agreement with the faintest observed SNe~Iax for deflagration models with remnant contribution included we require models with an ejecta component \tilda a magnitude or more fainter than the faintest model from the L22 sequence. A future investigation of whether it is possible to produce even fainter pure deflagration models, something L22 suggested can be achieved for single-spark deflagration models, would therefore be of interest.  

\subsection{Implications for the remnant}
\label{sec:remnant_implications}
While our models which produce good agreement with the bright SNe~Iax, SN~2005hk, adopt the full remnant \nickel~mass predicted by L22 our models in best agreement with the intermediate-luminosity SNe~Iax, SN~2019muj and faint SNe~Iax, SN~2020kyg, had remnant \nickel~masses one third and one tenth of those predicted by L22 respectively (see Table \ref{tab:remnant_properties}). From Table \ref{tab:model_properties} we see that moving to fainter deflagration models the remnant \nickel~mass predicted increases relative to the ejecta \nickel~mass. Adopting the entire remnant \nickel~mass predicted by L22 and assuming all energy emitted from the \nickel~decay chain in the remnant contributes to the optical display of the faintest L22 deflagration model (r114\_d6.0\_Z) produces an event of comparable brightness to SN~2019muj (see Section \ref{subsec:Intermediate_Iax_Lightcurves}). L22 suggest it should be possible to produce models fainter than r114\_d6.0\_Z. However, utilising the entire predicted remnant \nickel~mass in our intermediate-luminosity and faint deflagration models results in a very large change in the overall properties of the models leading to poor agreement with observed SNe~Iax. This suggests not all the energy injected in the remnant from radioactive decays can be used to power the light curves of intermediate-luminosity and faint models. This finding is consistent with what \cite{kromer2015a} found for their faint hybrid CONe WD pure deflagration model they compared to the faint SNe~Iax, SN~2008ha.

Models of SNe~Iax remnants by \cite{shen2017a} that predict some of the radioactive energy released powers a wind in the outer envelope of the remnant provide a potential explanation for this. If a large enough fraction of the energy released in the remnant is used to power such a wind, the energy contributing to the light curve will be significantly reduced. Alternatively, \cite{maeda2022a} have proposed a mechanism in which energy from radioactive decays causes hot \nickel~rich regions of the remnant to be ejected on the timescale of approximately a month after explosion, again a fraction of the energy injected in the remnant is required to power this process. As we do not take into account either of these mechanisms in our simulations these scenarios are roughly equivalent to having a reduced remnant \nickel~mass for our models (although time variations in the fraction of deposited  energy driving these mechanisms will not be captured). 

To estimate the potential impact of remnant winds, we calculated the approximate kinetic energy required to drive winds in the envelope of the remnant for the r48\_d5.0\_Z and r114\_d6.0\_Z models from the L22 sequence, using the SNe~Iax remnant wind models computed by \cite{shen2017a} as a guide. The \cite{shen2017a} models with greatest remnant masses have core and envelope masses of 0.9 and 0.1\,\msun~respectively, giving slightly lower but similar total remnant masses to the  L22 remnant masses predicted for the models used in this study (1.16 to 1.38\,\msun). Although we are unable to differentiate the remnant core and envelope from the L22 simulations we do have approximate yields of the IGEs and IMEs in the remnant which should primarily be found in the envelope and make up a sizeable fraction of the total envelope mass. The IGEs and IMEs predicted for the remnants sum to 0.074 and 0.053 \msun~for the r48\_d5.0\_Z and r114\_d6.0\_Z models respectively. As some of the envelope will be unburnt CO material, which we can not determine the mass of, an envelope mass of 0.1 \msun~seems reasonable for these models and we utilise this envelope mass as an initial guess for the mass driven by the radioactive wind in this calculation. The \cite{shen2017a} models have average winds speeds up to \tilda 1100 km s$^{-1}$ which we adopt for our calculation here to obtain an estimated upper limit on the energy used to drive winds in the remnant envelope.

The total energy deposited by the \nickel~decay chain over the simulation time is $4.85\times10^{48}$ erg for Model r48\_d5.0\_Z\_R\_6000K\_0.33Ni and $3.83\times10^{48}$ erg for Model r114\_d6.0\_Z\_R\_6000K\_0.1Ni. Models r48\_d5.0\_Z\_R\_6000K\_0.33Ni and r114\_d6.0\_Z\_R\_6000K\_0.1Ni require that two thirds and nine tenths of this energy contributes to processes other than powering the model light curves, respectively. From the calculation discussed above we estimate the total kinetic energy required to drive the remnant wind over the simulation time is $1.20\times10^{48}$ erg. For both models, this accounts for just over a third of the discrepancy between the energy we inject in the model remnants and the energy we predict would be deposited in the remnants from the L22 remnant \nickel~mass estimates. For envelope masses higher than the 0.1\,\msun~we estimated for these calculations an even greater proportion of the energy injected from radioactive decays in the remnant could be driving winds. It is therefore plausible that energy used to drive such winds can explain why only a fraction of the total energy injected in the remnants from radioactive decays is required to power our r48\_d5.0\_Z\_6000K\_0.33Ni and r114\_d6.0\_Z\_6000K\_0.1Ni model light curves. 

\cite{shen2017a} also suggest the hot relatively high density regime of the remnant may be so highly ionised that \nickel~and \cobalt~have radioactive decay rates lower than those we typically use in our simulations, which are appropriate for atomic or low ionisation species. This reduction in decay rate is expected to be relevant in highly ionised regimes with densities $\lesssim$ $10^5$ g cm$^{-3}$ (see \citealt{shen2017a} for details). The remnants predicted by the pure deflagration simulations of L22 have maximum central densities of \tilda $10^5$ g cm$^{-3}$. Therefore, it is plausible that regions of the remnant may inhabit this regime of reduced decay rates providing another possible explanation for our requirement to adopt lower remnant \nickel~masses than those predicted by L22.  

\cite{maeda2022a} alternatively suggest that diffusion timescales in the remnant are long enough that they lead to significant delays in the radiation contributing to the optical display of SNe~Iax. In their proposed scenario \nickel~rich material is ejected from the remnant around a month after explosion. They estimate diffusion in this material will become efficient by around 20 days since explosion allowing radiation originating there to begin to contribute to the optical display. However, they suggest it could be \tilda a year before radiation produced from the decay of \nickel~and \cobalt~in the remaining bound material can diffuse efficiently and thus contribute to the optical display of the SNe~Iax. As we assume no significant diffusion time in the remnant in our models, this scenario could also explain the reduced remnant \nickel~masses we must adopt for intermediate luminosity and faint models.

As discussed above, adopting reduced \nickel~masses in our simulations can provide a rough approximation of the behaviour expected in our models at early times if delayed radioactive decays \citep{shen2017a} or significant diffusion times in the remnant \citep{maeda2022a} are considered. However, relative to these cases, reducing the remnant \nickel~mass will underestimate the total radiation emitted by the remnant over time. Additionally, the contribution of the remnant to the optical display at later times will be significantly underestimated. Determining how these processes would influence the synthetic observables predicted by our simulations requires detailed modelling of the remnant structure and evolution, which is outside the scope of this study. However, we speculate that if we were to account for delayed radioactive decays or significant diffusion times in the remnant in our simulations, the remnant would likely have a reduced impact on the model light curves around peak but a greater influence on the light curve decline at later times relative to the simulations presented here. 

We finally note that it is also possible that the L22 simulations overestimate the remnant \nickel~masses. For fainter deflagration models, the mass of burnt material relative to the total WD mass is much smaller, meaning the yields of the burnt material are more uncertain for these models than for brighter more energetic models which burn more material. However, these uncertainties would have to account for very large reductions in the remnant \nickel~masses predicted by L22 for the models (up to a factor of 10) and it is therefore highly unlikely these uncertainties are relevant here.

\subsection{Future work}
\label{sec:future_work}
Including the energy injected from a luminous remnant in our radiative transfer simulations leads to significantly improved agreement with the spectra and light curves of observed SNe~Iax, strengthening the case of pure deflagration models for SNe~Iax. However, outstanding questions remain. Firstly, it is unclear how the assumptions of the radiative transfer impact our results. In this work, we have adopted the NLTE approximation described by \citet{kromer2009a}. However, including a full NLTE treatment of the plasma conditions has been shown to have important effects on the synthetic observables produced by radiative transfer simulations of SNe Ia, especially at later times \citep{blondin2013a, dessart2014a, shingles2020a, shingles2022a, shen2021a, collins2023a}. In future work we will utilise the full NLTE treatment of \cite{shingles2020a} to investigate the importance of NLTE effects. In particular, adopting the full NLTE treatment will enable us to extend our simulations to the late phases where the remnant contribution becomes more prominent as the ejecta become increasingly optically thin. This will allow us to investigate if the remnant is required to explain the peculiar late time spectra of observed SNe~Iax which never become truly nebular, as suggested by \cite{foley2016a}.

Important questions also remain around the nature of the remnant. Hydrodynamic explosion simulations of pure deflagrations following the remnants more closely would provide detailed remnant structures and compositions which would improve our understanding of the remnant structure providing insight into its likely evolution and emission. This is highly relevant to determining if the energy injection we utilise in this study is reasonable, particularly for our intermediate-luminosity and faint models that require remnant \nickel~masses lower than those predicted by L22. As such, it would also be useful to further investigate the possibility of radioactively driven winds and delayed radioactive decays as suggested by \cite{shen2017a} and the second ejecta component proposed to be ejected from the remnant envelope by \cite{maeda2022a}, in particular focusing on how introducing such mechanisms impacts the overall properties of pure deflagration models and their agreement with observed SNe~Iax. 

\section*{Acknowledgements}
FPC and SAS, acknowledge funding from STFC grant ST/X00094X/1.
This work used the DiRAC Data Intensive service (CSD3) at the University of Cambridge, managed by the University of Cambridge University Information Services on behalf of the STFC DiRAC HPC Facility (www.dirac.ac.uk). The DiRAC component of CSD3 at Cambridge was funded by BEIS, UKRI and STFC capital funding and STFC operations grants. DiRAC is part of the UKRI Digital Research Infrastructure. The authors gratefully acknowledge the Gauss Centre for Supercomputing e.V. (www.gauss-centre.eu) for funding this project by providing computing time on the GCS Supercomputer JUWELS at Jülich Supercomputing Centre (JSC).
CEC acknowledges funding by the European Union (ERC, HEAVYMETAL, 101071865). Views and opinions expressed are however those of the author(s) only and do not necessarily reflect those of the European Union or the European Research Council. Neither the European Union nor the granting authority can be held responsible for them. LJS acknowledges support by the European Research Council (ERC) under the European Union’s Horizon 2020 research and innovation program (ERC Advanced Grant KILONOVA No. 885281).
LJS acknowledges support by Deutsche Forschungsgemeinschaft (DFG, German Research Foundation) - Project-ID 279384907 - SFB 1245 and MA 4248/3-1. This work was supported in part by the European Union (ChETEC-INFRA, project no. 101008324), NSF/IReNA and Tschira Foundation. NumPy and SciPy \citep{oliphant2007a}, Matplotlib \citep{hunter2007a}  and \href{https://zenodo.org/records/8302355} {\textsc{artistools}}\footnote{\href{https://github.com/artis-mcrt/artistools/}{https://github.com/artis-mcrt/artistools/}} \citep{artistools2023a} were used for data processing and plotting.
\section*{Data Availability}
The light curves and spectra presented here are available on the Heidelberg
supernova model archive HESMA\footnote{\href{https://hesma.h-its.org}{https://hesma.h-its.org}}~\citep{kromer2017a}.



\bibliographystyle{mnras}
\bibliography{references} 



\appendix
\section{Light Curve parameter tables}

\begin{landscape}
\begin{table}
    \centering
    \begin{threeparttable}
        \begin{tabular}{cccccccccccccccc}
            \toprule
    Model & $t_{rise}^{bol}$  & $M_{peak}^{bol}$  & ${\Delta}m_{15}^{bol}$ & $t_{rise}^u$  & $M_{peak}^u$  & ${\Delta}m_{15}^u$ & $t_{rise}^g$  & $M_{peak}^g$  & ${\Delta}m_{15}^g$ \\  
    \midrule    
    r10\_d4.0\_Z &                   $11.52^{12.00}_{10.93}$ & $-17.35^{-17.65}_{-17.23}$ & $1.09^{1.26}_{0.97}$ & $9.30^{10.20}_{8.05}$ & $-16.70^{-17.24}_{-16.41}$ & $3.48^{3.95}_{3.01}$   & $11.08^{11.69}_{10.47}$ & $-17.49^{-17.68}_{-17.30}$ & $1.71^{1.83}_{1.60}$ \\ [0.2cm]
    r10\_d4.0\_Z\_R\_2000K &         $14.11^{14.76}_{13.69}$ & $-17.69^{-18.03}_{-17.52}$ & $0.77^{1.06}_{0.62}$ & $12.99^{13.80}_{12.32}$ & $-17.09^{-17.71}_{-16.72}$ & $3.20^{3.71}_{2.81}$ & $14.28^{14.93}_{13.50}$ & $-17.93^{-18.34}_{-17.65}$ & $1.58^{1.81}_{1.42}$ \\ [0.2cm]
    r10\_d4.0\_Z\_R\_8000K &         $14.08^{14.59}_{13.60}$ & $-17.70^{-18.04}_{-17.51}$ & $0.84^{1.07}_{0.66}$ & $12.99^{13.74}_{12.00}$ & $-17.10^{-17.72}_{-16.73}$ & $3.13^{3.53}_{2.84}$ & $14.21^{14.66}_{13.57}$ & $-17.94^{-18.34}_{-17.65}$ & $1.51^{1.65}_{1.39}$ \\ [0.2cm]
    r10\_d4.0\_Z\_R\_15000K &        $14.08^{14.59}_{13.63}$ & $-17.70^{-18.04}_{-17.52}$ & $0.84^{1.07}_{0.71}$ & $12.99^{13.57}_{11.94}$ & $-17.10^{-17.72}_{-16.73}$ & $3.09^{3.57}_{2.77}$ & $14.25^{14.69}_{13.74}$ & $-17.94^{-18.35}_{-17.68}$ & $1.47^{1.64}_{1.34}$ \\ [0.2cm]
    r10\_d4.0\_Z\_R\_const\_rad &    $14.08^{14.85}_{13.48}$ & $-17.70^{-18.05}_{-17.52}$ & $0.83^{1.07}_{0.70}$ & $12.99^{13.57}_{12.26}$ & $-17.10^{-17.74}_{-16.70}$ & $3.13^{3.57}_{2.78}$ & $14.25^{14.69}_{13.70}$ & $-17.93^{-18.34}_{-17.66}$ & $1.51^{1.65}_{1.39}$ \\ [0.2cm]    
    r48\_d5.0\_Z &                   $7.38^{8.29}_{6.70}$  & $-15.81^{-16.22}_{-15.59}$ & $1.34^{1.62}_{1.18}$   & $5.53^{6.21}_{4.49}$ & $-15.29^{-15.98}_{-14.81}$ & $4.27^{4.77}_{4.04}$    & $7.20^{7.75}_{6.56}$ & $-15.95^{-16.39}_{-15.69}$ & $1.96^{2.15}_{1.80}$ \\ [0.2cm]
    r48\_d5.0\_Z\_R\_6000K\_0.33Ni & $9.07^{9.46}_{8.63}$  & $-16.33^{-16.66}_{-16.09}$ & $1.07^{1.38}_{0.87}$   & $8.42^{8.89}_{8.03}$ & $-15.92^{-16.47}_{-15.52}$ & $4.01^{4.61}_{3.68}$    & $9.24^{9.52}_{8.87}$ & $-16.58^{-16.88}_{-16.32}$ & $1.79^{1.97}_{1.65}$ \\ [0.2cm]
    r114\_d6.0\_Z &                  $4.36^{4.72}_{4.04}$  & $-14.91^{-14.99}_{-14.79}$ & $2.43^{2.51}_{2.36}$   & $2.93^{3.64}_{2.50}$ & $-14.62^{-14.78}_{-14.37}$ & $4.85^{5.35}_{4.45}$    & $4.62^{4.81}_{4.33}$ & $-15.11^{-15.20}_{-15.01}$ & $2.30^{2.42}_{2.18}$ \\ [0.2cm]
    r114\_d6.0\_Z\_R\_4000K &        $6.94^{7.59}_{4.78}$  & $-16.74^{-16.89}_{-16.40}$ & $1.14^{1.23}_{0.88}$   & $6.08^{6.75}_{4.93}$ & $-16.78^{-16.92}_{-16.44}$ & $4.64^{5.07}_{4.37}$    & $7.64^{8.49}_{6.35}$ & $-16.52^{-16.61}_{-16.34}$ & $1.81^{1.95}_{1.67}$ \\ [0.2cm]
    r114\_d6.0\_Z\_R\_6000K\_0.1Ni & $5.09^{5.36}_{4.88}$  & $-15.27^{-15.35}_{-15.07}$ & $1.68^{1.78}_{1.59}$   & $4.12^{4.36}_{3.80}$ & $-15.02^{-15.16}_{-14.65}$ & $5.02^{5.39}_{4.72}$    & $5.61^{5.64}_{5.29}$ & $-15.53^{-15.61}_{-15.34}$ & $2.24^{2.32}_{2.11}$ \\ 
    \bottomrule

        \toprule
    Model & $t_{rise}^{r}$  & $M_{peak}^{r}$  & ${\Delta}m_{15}^{r}$ & $t_{rise}^i$  & $M_{peak}^i$  & ${\Delta}m_{15}^i$ & $t_{rise}^z$  & $M_{peak}^z$  & ${\Delta}m_{15}^z$ \\  
    
    \midrule
    r10\_d4.0\_Z &                   $14.05^{14.75}_{13.42}$ & $-17.86^{-17.97}_{-17.80}$ & $1.25^{1.41}_{1.13}$ & $16.24^{18.07}_{13.82}$ & $-17.47^{-17.53}_{-17.43}$ & $1.10^{1.28}_{0.92}$ & $18.03^{18.93}_{16.86}$ & $-17.71^{-17.79}_{-17.65}$ & $1.20^{1.33}_{1.09}$ \\ [0.2cm]
    r10\_d4.0\_Z\_R\_2000K &         $16.51^{17.68}_{15.42}$ & $-18.13^{-18.22}_{-18.07}$ & $0.94^{1.07}_{0.82}$ & $21.55^{23.50}_{15.06}$ & $-17.57^{-17.63}_{-17.51}$ & $0.86^{1.02}_{0.36}$ & $22.72^{23.85}_{21.51}$ & $-17.91^{-18.02}_{-17.85}$ & $1.00^{1.12}_{0.87}$ \\ [0.2cm]
    r10\_d4.0\_Z\_R\_8000K &         $16.78^{18.19}_{15.42}$ & $-18.14^{-18.23}_{-18.07}$ & $0.79^{0.89}_{0.70}$ & $21.97^{23.58}_{20.61}$ & $-17.62^{-17.68}_{-17.58}$ & $0.64^{0.78}_{0.52}$ & $22.91^{23.93}_{21.43}$ & $-17.93^{-17.99}_{-17.87}$ & $0.73^{0.87}_{0.58}$ \\ [0.2cm]
    r10\_d4.0\_Z\_R\_15000K &        $16.55^{17.95}_{15.61}$ & $-18.13^{-18.23}_{-18.07}$ & $0.72^{0.83}_{0.62}$ & $21.78^{23.97}_{16.63}$ & $-17.58^{-17.65}_{-17.52}$ & $0.62^{0.79}_{0.39}$ & $22.95^{23.85}_{21.74}$ & $-17.91^{-17.98}_{-17.86}$ & $0.69^{0.81}_{0.59}$ \\ [0.2cm]
    r10\_d4.0\_Z\_R\_const\_rad &    $16.70^{17.68}_{15.61}$ & $-18.14^{-18.24}_{-18.08}$ & $0.77^{0.90}_{0.68}$ & $21.97^{23.54}_{20.33}$ & $-17.61^{-17.76}_{-17.56}$ & $0.63^{0.79}_{0.51}$ & $22.87^{23.93}_{21.55}$ & $-17.93^{-18.03}_{-17.86}$ & $0.72^{0.90}_{0.58}$ \\ [0.2cm]    
    r48\_d5.0\_Z &                   $10.16^{10.95}_{9.11}$ & $-16.24^{-16.41}_{-16.10}$ & $1.50^{1.67}_{1.37}$  & $11.83^{13.16}_{10.43}$ & $-15.97^{-16.09}_{-15.87}$ & $1.38^{1.49}_{1.23}$ & $12.17^{12.95}_{11.56}$ & $-16.12^{-16.29}_{-16.01}$ & $1.24^{1.35}_{1.09}$ \\ [0.2cm]
    r48\_d5.0\_Z\_R\_6000K\_0.33Ni & $12.85^{14.59}_{9.82}$ & $-16.58^{-16.72}_{-16.47}$ & $1.05^{1.22}_{0.86}$  & $15.64^{16.80}_{14.28}$ & $-16.20^{-16.30}_{-16.14}$ & $0.79^{0.90}_{0.68}$ & $15.95^{16.83}_{15.00}$ & $-16.42^{-16.54}_{-16.36}$ & $0.83^{0.97}_{0.65}$ \\ [0.2cm]    
    r114\_d6.0\_Z &                  $6.15^{6.39}_{5.64}$ & $-15.24^{-15.30}_{-15.17}$ & $2.73^{2.83}_{2.55}$    & $6.92^{7.60}_{6.11}$    & $-14.93^{-15.00}_{-14.84}$ & $2.78^{2.96}_{2.59}$ & $7.47^{7.75}_{7.07}$  & $-15.09^{-15.15}_{-14.96}$ & $2.00^{2.14}_{1.86}$ \\ [0.2cm]
    r114\_d6.0\_Z\_R\_4000K &        $8.26^{8.70}_{7.24}$ & $-16.41^{-16.56}_{-16.14}$ & $0.99^{1.12}_{0.84}$    & $9.75^{11.52}_{8.53}$   & $-16.37^{-16.61}_{-15.84}$ & $0.65^{0.82}_{0.44}$ & $10.78^{14.15}_{8.67}$ & $-16.58^{-16.82}_{-15.99}$ & $0.63^{0.90}_{0.31}$ \\ [0.2cm]    
    r114\_d6.0\_Z\_R\_6000K\_0.1Ni & $7.23^{7.66}_{6.79}$ & $-15.52^{-15.59}_{-15.40}$ & $1.66^{1.89}_{1.52}$    & $8.88^{9.64}_{8.25}$    & $-15.13^{-15.22}_{-14.99}$ & $1.37^{1.50}_{1.27}$ & $9.41^{10.17}_{8.91}$  & $-15.36^{-15.45}_{-15.15}$ & $1.35^{1.48}_{1.26}$ \\ 
    \bottomrule    
        \end{tabular}        
        \caption{Bolometric and \textit{ugriz} Sloan band light curve properties for our new models and the L22 models we compare to in this work. For each quantity the value determined from the angle-averaged light curve is quoted along with the range obtained from the model line-of-site dependent light curves (displayed as a ± range on each angle averaged value).}
        \label{tab:Bol_Sloan_lightcurve_properties}
    \end{threeparttable}
\end{table}
\end{landscape}

\begin{landscape}
\begin{table}
    \centering
    \begin{threeparttable}
        \begin{tabular}{cccccccccccccccc}
            \toprule
    Model & $t_{rise}^U$  & $M_{peak}^U$  & ${\Delta}m_{15}^U$ &
    $t_{rise}^B$  & $M_{peak}^B$  & ${\Delta}m_{15}^B$ &  
    $t_{rise}^V$  & $M_{peak}^V$  & ${\Delta}m_{15}^V$ &  
    $t_{rise}^R$  & $M_{peak}^R$  & ${\Delta}m_{15}^R$ \\  
    \midrule    
    r10\_d4.0\_Z                      & $9.42^{10.14}_{8.34}$   & $-17.57^{-18.09}_{-17.30}$ & $3.38^{3.71}_{3.00}$  & $10.40^{11.15}_{9.65}$  & $-17.28^{-17.70}_{-17.07}$ & $2.07^{2.21}_{1.96}$ & $12.85^{13.46}_{12.18}$ & $-17.79^{-18.00}_{-17.68}$ & $1.34^{1.45}_{1.24}$ & $14.09^{14.67}_{13.46}$ & $-17.94^{-18.04}_{-17.89}$ & $1.18^{1.34}_{1.08}$ \\ [0.2cm]
    r10\_d4.0\_Z\_R\_2000K            & $13.08^{13.86}_{12.44}$ & $-17.98^{-18.57}_{-17.61}$ & $3.12^{3.51}_{2.82}$  & $13.77^{14.55}_{12.95}$ & $-17.72^{-18.20}_{-17.38}$ & $1.93^{2.16}_{1.73}$ & $15.54^{16.08}_{14.87}$ & $-18.20^{-18.41}_{-18.07}$ & $1.18^{1.31}_{1.04}$ & $16.51^{17.64}_{15.38}$ & $-18.18^{-18.26}_{-18.13}$ & $0.87^{0.98}_{0.76}$ \\ [0.2cm]
    r10\_d4.0\_Z\_R\_8000K            & $13.08^{13.69}_{12.06}$ & $-17.99^{-18.59}_{-17.61}$ & $3.04^{3.36}_{2.75}$  & $13.70^{14.28}_{13.06}$ & $-17.73^{-18.20}_{-17.39}$ & $1.85^{2.00}_{1.68}$ & $15.57^{16.04}_{15.06}$ & $-18.20^{-18.42}_{-18.05}$ & $1.06^{1.17}_{0.94}$ & $16.86^{18.34}_{15.38}$ & $-18.19^{-18.27}_{-18.13}$ & $0.72^{0.83}_{0.64}$ \\ [0.2cm]
    r10\_d4.0\_Z\_R\_15000K           & $13.08^{13.63}_{12.06}$ & $-17.98^{-18.58}_{-17.62}$ & $3.00^{3.38}_{2.73}$  & $13.74^{14.25}_{12.99}$ & $-17.72^{-18.20}_{-17.39}$ & $1.81^{1.97}_{1.65}$ & $15.47^{16.31}_{14.95}$ & $-18.20^{-18.43}_{-18.08}$ & $0.99^{1.13}_{0.89}$ & $16.51^{18.07}_{15.49}$ & $-18.18^{-18.27}_{-18.13}$ & $0.66^{0.76}_{0.56}$ \\ [0.2cm]
    r10\_d4.0\_Z\_R\_const\_rad       & $13.11^{13.66}_{12.41}$ & $-17.99^{-18.60}_{-17.58}$ & $3.05^{3.30}_{2.76}$  & $13.74^{14.28}_{13.09}$ & $-17.73^{-18.21}_{-17.39}$ & $1.85^{2.04}_{1.71}$ & $15.54^{16.04}_{14.91}$ & $-18.20^{-18.43}_{-18.07}$ & $1.05^{1.17}_{0.87}$ & $16.74^{18.03}_{15.49}$ & $-18.19^{-18.28}_{-18.14}$ & $0.70^{0.82}_{0.62}$ \\ [0.2cm]
    r48\_d5.0\_Z                      & $5.63^{6.28}_{4.98}$    & $-16.14^{-16.80}_{-15.68}$ & $4.17^{4.59}_{3.96}$  & $6.69^{7.24}_{6.11}$    & $-15.76^{-16.27}_{-15.45}$ & $2.39^{2.59}_{2.21}$  & $8.46^{9.28}_{7.51}$   & $-16.17^{-16.44}_{-16.01}$ & $1.55^{1.68}_{1.43}$ & $10.43^{11.18}_{9.24}$  & $-16.34^{-16.51}_{-16.21}$ & $1.46^{1.60}_{1.33}$ \\ [0.2cm]
    r48\_d5.0\_Z\_R\_6000K\_0.33Ni    & $8.47^{8.84}_{8.13}$    & $-16.79^{-17.30}_{-16.40}$ & $3.89^{4.31}_{3.58}$  & $8.94^{9.35}_{8.60}$    & $-16.42^{-16.76}_{-16.12}$ & $2.22^{2.39}_{2.05}$  & $9.86^{10.67}_{9.31}$  & $-16.68^{-16.87}_{-16.52}$ & $1.22^{1.37}_{1.11}$ & $13.43^{14.83}_{9.92}$  & $-16.66^{-16.79}_{-16.56}$ & $0.99^{1.13}_{0.82}$ \\ [0.2cm]
    r114\_d6.0\_Z                     & $3.12^{3.61}_{2.72}$    & $-15.45^{-15.58}_{-15.21}$ & $4.74^{5.13}_{4.38}$  & $4.26^{4.47}_{4.07}$    & $-14.98^{-15.07}_{-14.84}$ & $2.81^{2.94}_{2.69}$  & $5.45^{5.86}_{5.11}$   & $-15.22^{-15.28}_{-15.14}$ & $2.16^{2.26}_{2.06}$ & $6.15^{6.42}_{5.61}$    & $-15.34^{-15.39}_{-15.27}$ & $2.71^{2.80}_{2.55}$ \\ [0.2cm]
    r114\_d6.0\_Z\_R\_4000K           & $6.08^{6.73}_{5.09}$    & $-17.57^{-17.71}_{-17.24}$ & $4.46^{4.88}_{4.21}$  & $7.24^{7.95}_{5.50}$    & $-16.51^{-16.62}_{-16.33}$ & $2.37^{2.55}_{2.19}$  & $7.71^{8.43}_{6.90}$   & $-16.52^{-16.62}_{-16.35}$ & $1.32^{1.49}_{1.17}$ & $8.56^{9.04}_{7.44}$    & $-16.55^{-16.72}_{-16.24}$ & $0.90^{1.01}_{0.75}$ \\ [0.2cm]
    r114\_d6.0\_Z\_R\_6000K\_0.1Ni    & $4.26^{4.50}_{3.93}$    & $-15.86^{-15.99}_{-15.50}$ & $4.91^{5.25}_{4.62}$  & $5.23^{5.45}_{4.99}$    & $-15.40^{-15.49}_{-15.16}$ & $2.79^{2.91}_{2.65}$  & $6.45^{6.70}_{6.23}$   & $-15.58^{-15.67}_{-15.45}$ & $1.78^{1.90}_{1.67}$ & $7.32^{7.75}_{6.88}$    & $-15.60^{-15.65}_{-15.48}$ & $1.58^{1.78}_{1.47}$ \\
    \bottomrule

        \toprule
    Model & $t_{rise}^I$  & $M_{peak}^I$  & ${\Delta}m_{15}^I$ & 
    $t_{rise}^J$  & $M_{peak}^J$  & ${\Delta}m_{15}^J$ &
    $t_{rise}^H$  & $M_{peak}^H$  & ${\Delta}m_{15}^H$ &
    $t_{rise}^K$  & $M_{peak}^K$  & ${\Delta}m_{15}^K$ \\
    
    \midrule
    r10\_d4.0\_Z                        & $16.86^{18.15}_{15.34}$ & $-18.04^{-18.11}_{-17.99}$ & $1.23^{1.37}_{1.10}$ & $11.49$ & $-17.14$ & $0.68$ & $21.22$ & $-17.58$ & $1.10$ & $21.00$ & $-17.45$ & $1.25$ \\ [0.2cm]
    r10\_d4.0\_Z\_R\_2000K              & $21.74^{23.30}_{20.57}$ & $-18.19^{-18.27}_{-18.12}$ & $0.98^{1.11}_{0.89}$ &    -   &      -    &   -    &    -    &     -    &   -    &    -     &     -    &   -    \\ [0.2cm]
    r10\_d4.0\_Z\_R\_8000K              & $21.97^{23.26}_{20.76}$ & $-18.22^{-18.29}_{-18.17}$ & $0.74^{0.88}_{0.64}$ & $10.96$ & $-17.21$ & $0.23$ & $26.77$ & $-17.78$ & $0.54$ & $27.21$  & $-17.58$ & $0.52$ \\ [0.2cm]
    r10\_d4.0\_Z\_R\_15000K             & $21.86^{23.30}_{20.22}$ & $-18.20^{-18.28}_{-18.14}$ & $0.70^{0.87}_{0.62}$ & $11.05$ & $-17.20$ & $0.40$ & $26.11$ & $-17.69$ & $0.59$ & $26.07$  & $-17.48$ & $0.65$ \\ [0.2cm]
    r10\_d4.0\_Z\_R\_const\_rad         & $21.97^{23.73}_{20.73}$ & $-18.22^{-18.30}_{-18.17}$ & $0.73^{0.87}_{0.61}$ & $10.96$ & $-17.20$ & $0.23$ & $26.69$ & $-17.77$ & $0.51$ & $27.70$  & $-17.57$ & $0.47$ \\ [0.2cm]
    r48\_d5.0\_Z                        & $11.80^{12.51}_{11.12}$ & $-16.51^{-16.65}_{-16.40}$ & $1.41^{1.52}_{1.30}$ & $10.30$ & $-15.68$ & $0.93$ & $13.96$ & $-16.00$ & $1.21$ & $14.09$  & $-15.91$ & $1.59$ \\ [0.2cm]
    r48\_d5.0\_Z\_R\_6000K\_0.33Ni      & $15.51^{16.39}_{14.72}$ & $-16.77^{-16.87}_{-16.68}$ & $0.89^{1.01}_{0.78}$ & $14.09$ & $-15.83$ & $0.21$ & $21.31$ & $-16.33$ & $0.51$ & $22.94$  & $-16.26$ & $0.50$ \\ [0.2cm]
    r114\_d6.0\_Z                       & $7.11^{7.50}_{6.70}$    & $-15.51^{-15.58}_{-15.40}$ & $2.50^{2.63}_{2.34}$ & $4.76$  & $-14.69$ & $2.17$ & $9.07$ & $-14.98$  & $3.09$ & $8.90$   & $-14.97$ & $3.78$ \\ [0.2cm]
    r114\_d6.0\_Z\_R\_4000K             & $9.99^{12.51}_{8.90}$   & $-16.89^{-17.16}_{-16.37}$ & $0.67^{0.83}_{0.42}$ & $11.67$ & $-17.49$ & $0.56$ & $12.19$ & $-17.9$  & $0.63$ & $14.09$  & $-18.03$ & $0.52$ \\ [0.2cm]
    r114\_d6.0\_Z\_R\_6000K\_0.1Ni      & $9.06^{9.74}_{8.25}$    & $-15.73^{-15.81}_{-15.55}$ & $1.48^{1.62}_{1.40}$ & $5.54$  & $-14.81$ & $0.46$ & $11.49$ & $-15.56$ & $1.18$ & $11.90$  & $-15.42$ & $1.13$ \\ 
    \bottomrule    
        \end{tabular}        
        \caption{Same as Table~\ref{tab:Bol_Sloan_lightcurve_properties} for the Bessel \textit{UBVRIJHK} bands. Note, the lower relative flux of the models in the \textit{JHK}-bands meant reliable values could not be determined for the line-of-sight dependent light curve quantities due to the significant Monte Carlo noise. Therefore only the angle averaged values are quoted for these bands. Additionally, we do not include any \textit{JHK}-band light curve quantities for Model r10\_d4.0\_Z\_R\_2000K due to its peculiar light curve behaviour in these bands which we believe to be unphysical.} 
        \label{tab:Bessel_lightcurve_properties}
    \end{threeparttable}
\end{table}
\end{landscape}


\bsp	
\label{lastpage}
\end{document}